\documentclass[prd,onecolumn,showpacs,floatfix,superscriptaddress,nofootinbib]{revtex4-2}
\usepackage{graphicx}
\usepackage{epsfig}
\usepackage{bm}
\usepackage{amssymb}
\usepackage{float}
\usepackage{amsmath}
\usepackage{dcolumn}
\usepackage{cancel}

\usepackage{mathrsfs}
\usepackage{dcolumn}
\usepackage{graphicx}
\usepackage{amsmath}
\usepackage{amsfonts}
\usepackage{amssymb}
\usepackage{microtype}
\usepackage{subfigure}
\usepackage{makeidx}
\usepackage{bm}
\usepackage{epsf}
\usepackage{color}
\usepackage{multirow,dcolumn}
\usepackage{graphicx}
\usepackage{mathrsfs}
\graphicspath{{Images/}}

\def\doi{http://doi.org}

\def\be{\begin{equation*}}
\def\ee{\end{equation*}}

\def\Ref{\ref}

\begin{document}

\title{Magnetically Charged Euler-Heisenberg Black Holes with Scalar Hair}

\author{Thanasis Karakasis}
\email{thanasiskarakasis@mail.ntua.gr}
\affiliation{Physics Department, National Technical University of Athens, 15780 Zografou Campus, Athens, Greece}
	
	\author{George Koutsoumbas}
	\email{kutsubas@central.ntua.gr}
	\affiliation{Physics Department, National Technical University of Athens, 15780 Zografou Campus, Athens, Greece}

\author{Andri Machattou}
	\email{andrimachattou@hotmail.com}
	\affiliation{Physics Department, National Technical University of Athens, 15780 Zografou Campus, Athens, Greece}
	
	\author{Eleftherios Papantonopoulos}
	\email{lpapa@central.ntua.gr} \affiliation{Physics Department, National Technical University of Athens, 15780 Zografou Campus, Athens, Greece}

\vspace{4.5cm}

\begin{abstract}
We study the Einstein-Euler-Heisenberg theory in the presence of a  self interacting scalar field, minimally coupled to gravity. We solve analytically the field equations for the  magnetically charged case and we obtain novel magnetically charged hairy black holes. The scalar field dresses the black hole with a secondary scalar hair. The hairy black hole develops three horizons when Euler-Heisenberg parameter and the magnetic charge are  small and the horizon radius is getting large when the scalar charge and the gravitational mass are large. The presence of matter and the magnetic field outside the horizon of the black hole increases the temperature only for small black holes. Calculating the heat capacity we show that  the asymptotically AdS Euler-Heisenberg hairy black hole undergoes a second order phase transition and then it is stabilized. Also the weak energy condition is violated for the asymptotically AdS Euler-Heisenberg hairy black hole.
\end{abstract}

\maketitle

\flushbottom

\tableofcontents

\section{Introduction}
The no-hair theorem states that black holes are described via charges that can be measured at infinity by an observer, and as a result there can be no additional degrees of freedom to characterize the black hole geometry. Hence, black holes are described by their mass, angular momentum and electromagnetic charges. The no-hair conjecture implies that a  black hole spacetime does not allow degrees of freedom that are responsible for the formation of primary and secondary hair. The first ones are not associated to the conserved black hole parameters mentioned previously, while the second ones concern non-trivial fields that interact with the black hole parameters. The no-hair theorem affects all types of fields, scalar, gauge, tensor fields etc. We will only discuss the no-scalar hair theorem. The first proof of the no-scalar hair theorem was given by Bekenstein \cite{Bekenstein:1972ny, Bekenstein:1972ky}. The main result is that a positive scalar potential such as a mass term potential cannot support a hairy black hole structure. As a result, if one wants to violate the no-hair theorem, a negative potential has to be considered for the scalar field \cite{Herdeiro:2015waa}.

The first hairy black hole which was an exact black hole solution with scalar field, called BBMB black hole, was found by Bocharova, Bronnikov and Melnikov and independently by Bekenstein \cite{BBMB}. However, it was found that the scalar field was divergent at the horizon and also the stability analysis showed that they were unstable \cite{bronnikov}, so these solutions were not physically acceptable.
The action of the BBMB black hole consists of the Ricci scalar, a scalar field with a  kinetic term and a conformal coupling between matter and curvature. To cure the horizon-divergence problem, a cosmological constant was considered which is  introducing a  length scale, and  a self interacting potential for the scalar field was  considered at first \cite{Martinez:2002ru} and exact dS black holes were obtained with a regular scalar field at the horizon. However, the solutions turned out to be unstable again \cite{Harper:2003wt}. In \cite{Karakasis:2021rpn} a scale in the form of non-linear modification of gravity was introduced and the scalar field has been shown to be finite at the horizon. All these solutions consider a theory that is conformally invariant and there is  a non-minimal coupling between matter and gravity.

Considering now the minimal coupling case, the first exact black hole solution was presented in \cite{Martinez:2004nb}, the Martinez, Troncoso, Zanelli (MTZ) black hole. The scalar potential is fixed $ad~hoc$, the geometry of the solution is hyperbolic and the scalar field remains finite at the black hole event horizon. This solution was generalized with the presence of an electromagnetic field \cite{Martinez:2006an}. In \cite{Kolyvaris:2009pc}, a potential that breaks the conformal invariance of the action of the MTZ black hole in the Jordan frame was considered and new black hole solutions where derived. In \cite{Gonzalez:2013aca} the scalar field was fixed $ad~hoc$ and novel black hole solutions were investigated, letting the scalar potential to be determined from the equations. The mass term comes as an independent integration constant and, as a result, the length scale introduced by the scalar field is a primary scalar hair. In \cite{Gonzalez:2014tga} the electrically charged case was considered and in \cite{Barrientos:2016ubi} the solution was generalized in the presence of non-linear electrodynamics. In this case the mass has contributions from  both an integration constant and the scalar length scale, so the scalar field dresses the black hole with secondary hair. Static black holes with a charged scalar field and conformally invariant Maxwell electrodynamics in $D$-dimensions have also been obtained numerically in \cite{Rahmani:2020vvv}. The shadow of hairy rotating black holes in view of the results of the Event Horizon Telescope were discussed in \cite{Khodadi:2021gbc} and in \cite{Konoplya:2021slg} the shadow of slowly rotating Gauss Bonnet with a dilaton field were investigated.

One can introduce several couplings between matter and curvature to evade the no-hair theorem, or consider modified theories of gravity. Couplings motivated from cosmology such as the non-minimal derivative coupling, i.e. a coupling between the Einstein tensor and the kinetic term of the scalar field have been investigated in \cite{Anabalon:2012tu,Kolyvaris:2013zfa,Kolyvaris:2011fk,Rinaldi:2012vy,Papantonopoulos:2019eff,Minamitsuji:2013ura,Anabalon:2013oea,Cisterna:2014nua}. These gravity theories are generated from  the Horndeski class Lagrangian, leading to second order differential equations for the metric fields \cite{Horndeski:1974wa}. More black hole solutions regarding Horndeski Lagrangian can be found in refs. \cite{Babichev:2013cya,Babichev:2017guv,Charmousis:2014zaa,Babichev:2015rva,Bravo-Gaete:2013dca,Bravo-Gaete:2014haa}. The Gauss-Bonnet invariant coupled to the scalar field has been also considered where it was shown that the no-hair theorem can be evaded and  asymptotically flat and AdS black holes carrying a scalar charge have been obtained \cite{Kanti:1995vq,Sotiriou:2013qea,Sotiriou:2014pfa,Antoniou:2017acq,Doneva:2017bvd,Silva:2017uqg,Doneva:2018rou} $f(R)$ gravity black holes have also been discussed \cite{Tang:2020sjs,Karakasis:2021lnq,Karakasis:2021ttn}. Couplings of the scalar field with electromagnetism has also been used to construct hairy black holes \cite{Garfinkle:1990qj,Anabalon:2013qua,Astefanesei:2019mds,Gibbons:1982ih,Priyadarshinee:2021rch,Yu:2020rqi}.

The Euler-Heisenberg Lagrangian of electrodynamics was at first considered in 1936 \cite{Heisenberg:1936nmg}. The Euler-Heinsenberg theory is a more accurate classical approximation of QED than Maxwell's theory, when the fields have high intensity. The vacuum is treated as a specific type of medium, and the properties of polarization and magnetization are determined by the clouds of virtual charges surrounding the real currents and charges \cite{Obukhov:2002xa}. A way to detect the effect of the Euler-Heisenberg theory has been proposed in \cite{Brodin:2001zz}. Since the Euler-Heisenberg theory has interesting physical features, it was a natural consequence to couple the Euler-Heineberg Lagrangian to the Ricci scalar via the volume element to search for black hole solutions. The first black hole solution to the Euler-Heisenberg electrodynamics was derived in \cite{Yajima:2000kw}, where analytical solutions were obtained for the magnetically charged case, also discussing electric charges and dyons. Electrically charged black holes were considered in \cite{Ruffini:2013hia} and \cite{Amaro:2020xro}, while in \cite{Amaro:2020xro} the geodesic structure was the main study of the paper. In \cite{Chen:2022tbb} motions of charged particles around the Euler-Heisenberg AdS black hole were studied. The thermodynamics of these black holes were studied in \cite{Magos:2020ykt, Dai:2022mko}, while the quasinormal modes were calculated in \cite{Breton:2021mju}. Rotating black holes were found in \cite{Breton:2019arv, Breton:2022fch}, while the Euler-Heisenberg Lagrangian was introduced along with modified gravity theories in \cite{Stefanov:2007bn, Guerrero:2020uhn, Nashed:2021ctg} and the corresponding black holes were analyzed. Finally, the shadow of Euler-Heisenberg black holes was investigated in \cite{Allahyari:2019jqz}.

The Euler-Heisenberg theory describes accurately the behaviour of electric and magnetic charges. However,  in the presence of a strong gravitational field in a astrophysical environment, for example near the horizon of a astrophysical black hole, the electric charge is considered to be negligible due to the presence of plasma (electrically charged particles result in electrically conductive plasma) around astrophysical black holes that neutralizes any electric charge carried by the black hole. For the upper bounds of electric charge an astrophysical black hole can carry we refer to \cite{Zajacek:2018ycb}. On the other hand, magnetically charged black holes cannot be neutralized with ordinary matter \cite{Maldacena:2020skw}. Recent  astrophysical observations showed that  many strong radio sources take the form of two emitting regions situated on opposite sites of a galaxy. To explain these astrophysical results a  theory was proposed  that the magnetic fields and high energy particles responsible for the synchrotron radiation were blown out of the galactic halo  in a giant explosion. Then, a possible explanation was proposed that such explosions could be generated from gravitational collapse and a model was used presented in \cite{Melvin:1963qx} which it was composed by a configuration that contains only electromagnetic field in a form of a collection of parallel magnetic lines  which is held together by its own gravitational attraction known as a static Melvin Universe. This theory is described  as a rigorous static cylindrically-symmetric solution of the combined sourceless Einstein-Maxwell equations \cite{Melvin:1965}-\cite{Thorne2}.

From the above discussion it would be very helpful to try to understand the behaviour of a  magnetic field in the presence of matter. The Euler-Heisenberg theory can provide a very reliable framework to study this effect. Motivated by this fact, in this work, we generalize the Einstein-Euler-Heisenberg black holes of \cite{Yajima:2000kw} by introducing matter parameterized by a self interacting scalar field, minimally coupled to gravity. By assuming only magnetic charges, we integrate analytically the field equations and discuss the corresponding solutions.
We found that when the Euler-Heisenberg parameter vanishes we obtain novel magnetically charged hairy black holes, while when the scalar charge vanish, we get the solution of \cite{Yajima:2000kw} while when both the  Euler-Heisenberg parameter and the magnetic charge vanishes we go back to the well known hairy black hole solution of \cite{Gonzalez:2013aca}. The scalar field dresses the black hole with  secondary scalar hair, since the scalar charge is related to the mass parameter, while the scalar potential is negative in order to support the hairy structure and it possesses a mass term that satisfies the Breitenlohner-Friedman bound that ensures the perturbative stability of the AdS spacetime.

The black hole horizon shrinks as the magnitude of the scalar field is getting larger, while is getting larger as the gravitational mass is increasing. Calculating thermodynamical quantities we found that the presence of matter outside the horizon of the black hole and also the presence of the magnetic field increases the temperature  for small black holes. We also found that  the temperature develops a minima in the AdS case signalizing in this way a second order phase transition, while the scalar field gains entropy for the black hole by the addition of a linear term in the entropy and hence the hairy black holes are thermodynamically preferred. Calculating the weak energy condition we find that it is violated in the case of asymptotically AdS spacetime.

The work is organized as follows. In Section \ref{sec1} we set up the theory, derive the solution and discuss the effect of the scalar field on the black hole. In Section \ref{sec2} we write down some limiting behaviors of the obtained black hole solution. In Section \ref{sec3} we discuss the thermodynamical properties, while in Section \ref{sec4} we investigate the energy conditions and finally in Section \ref{sec5} we conclude.

\section{Black hole solutions}
\label{sec1}

We consider the Euler-Heisenberg action in the presence of a scalar field
\begin{equation} S=\int d^4x\sqrt{-g} \mathcal{L} = \int d^4x\sqrt{-g}\left(\frac{R}{2} - \frac{1}{2}\partial^{\mu}\phi\partial_{\mu}\phi - V(\phi) -P+\alpha P^2 +\beta Q^2\right)~,
\label{action}
\end{equation}
where $\mathcal{L}$ denotes the Lagrangian of the theory, $P=F_{\mu\nu}F^{\mu\nu},~ Q = \epsilon_{\mu\nu\rho\sigma}F^{\mu\nu}F^{\rho\sigma}$, $F_{\mu\nu} = \partial_{\mu}A_{\nu} - \partial_{\nu}A_{\mu}$ is the Faraday tensor (field strength) and $\epsilon_{\mu\nu\rho\sigma}$ is the Levi-Civita tensor that satisfies
\begin{equation}\epsilon_{\mu\nu\rho\sigma}\epsilon^{\mu\nu\rho\sigma} =-24~.\end{equation}
The field equations are
\begin{eqnarray}
&&G_{\mu\nu} = T_{\mu\nu} \equiv T_{\mu\nu}^{\phi}+ T_{\mu\nu}^{EM}~,\label{emt}\\
&&\Box \phi = \frac{dV}{d\phi}~, \label{KG}\\
&&\nabla_{\mu}\left(F^{\mu\nu}-2\alpha PF^{\mu\nu}-2\beta Q\epsilon^{\mu\nu\xi\eta}F_{\xi\eta}\right)=0~,\\
&&T_{\mu\nu}^{\phi} = \partial_{\mu}\phi\partial_{\nu}\phi - \frac{1}{2}g_{\mu\nu}\partial^{\alpha}\phi\partial_{\alpha}\phi - g_{\mu\nu}V(\phi)~,\\
&&T_{\mu\nu}^{EM} = 2F_{\mu\rho}F_{\nu}^{~\rho} + \frac{1}{2}g_{\mu\nu}\left(-P+\alpha P^2+\beta Q^2\right) - 4\alpha P F_{\mu\rho}F_{~\nu}^{\rho} -8\beta Q\epsilon_{\mu\zeta\eta\rho}F^{\zeta\eta}F^{\rho}_{~\nu}~.\label{enmom}
\end{eqnarray}
We consider the following spherically symmetric ansatz for the spacetime metric
\begin{equation}
ds^2 = -b(r)dt^2 +b(r)^{-1}dr^2 +b_1(r)^2 d\Omega^2~,
\end{equation}
where $d\Omega^2 = d\theta^2 + \sin^2 \theta d\varphi^2$ which allows us to consider the following electromagnetic ansatz for the four-vector $A_{\mu}$
\begin{equation} A_{\mu} = \left(\mathcal{A}(r),0,0,Q_m\cos\theta\right)~, \end{equation}
where $Q_m$ is the magnetic charge of the black hole and the magnetic part of the four vector will be null at the equatorial plane. Under these ansaetze, the scalar quantities $P,Q$ that enter the field equations read
\begin{eqnarray}
&&P = \frac{2 Q_m^2}{b_1(r){}^4}-2 \mathcal{A}'(r)^2~,\\
&&Q = -\frac{8 Q_m \mathcal{A}'(r)}{b_1(r){}^2}~,
\end{eqnarray}
where it is clear that $Q$ will vanish if we do not consider dyons (both electric and magnetic charges).

\begin{figure}[h]
\centering
\includegraphics[width=.40\textwidth]{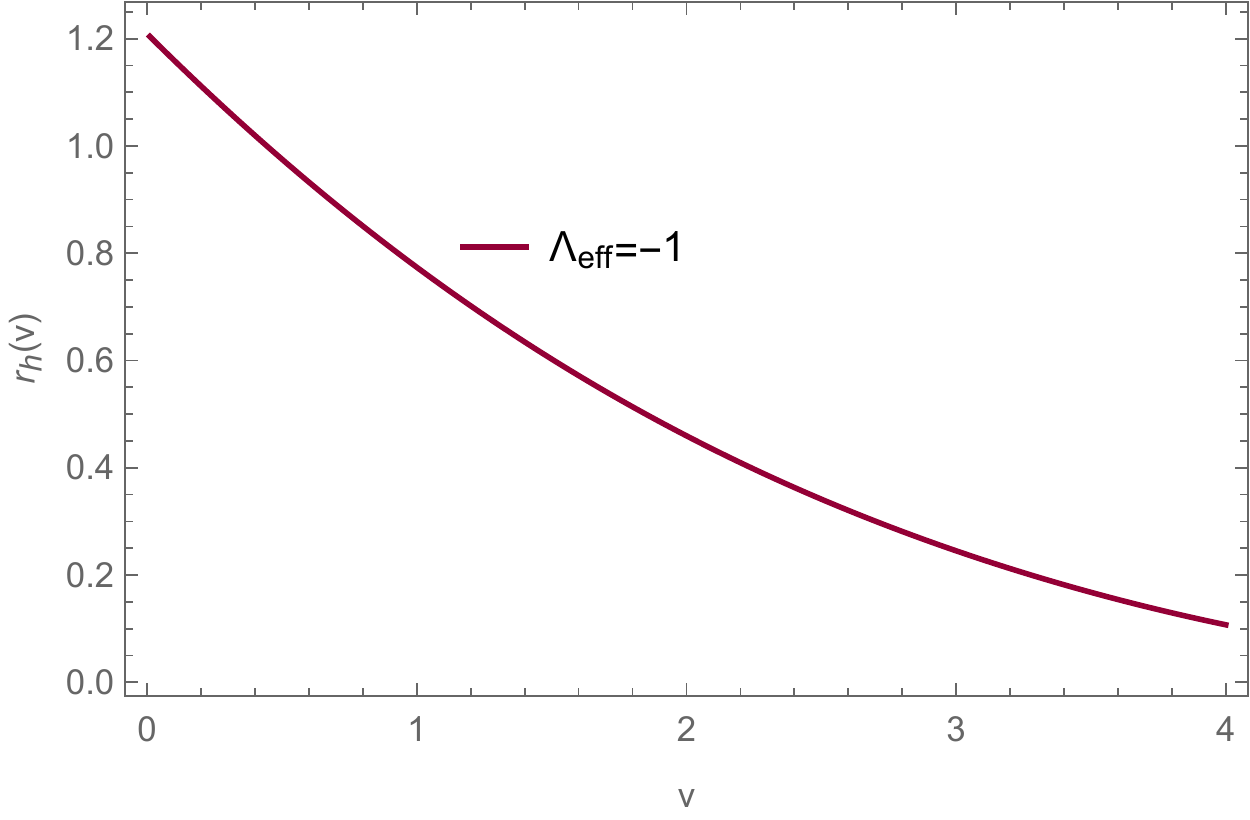}
\includegraphics[width=.40\textwidth]{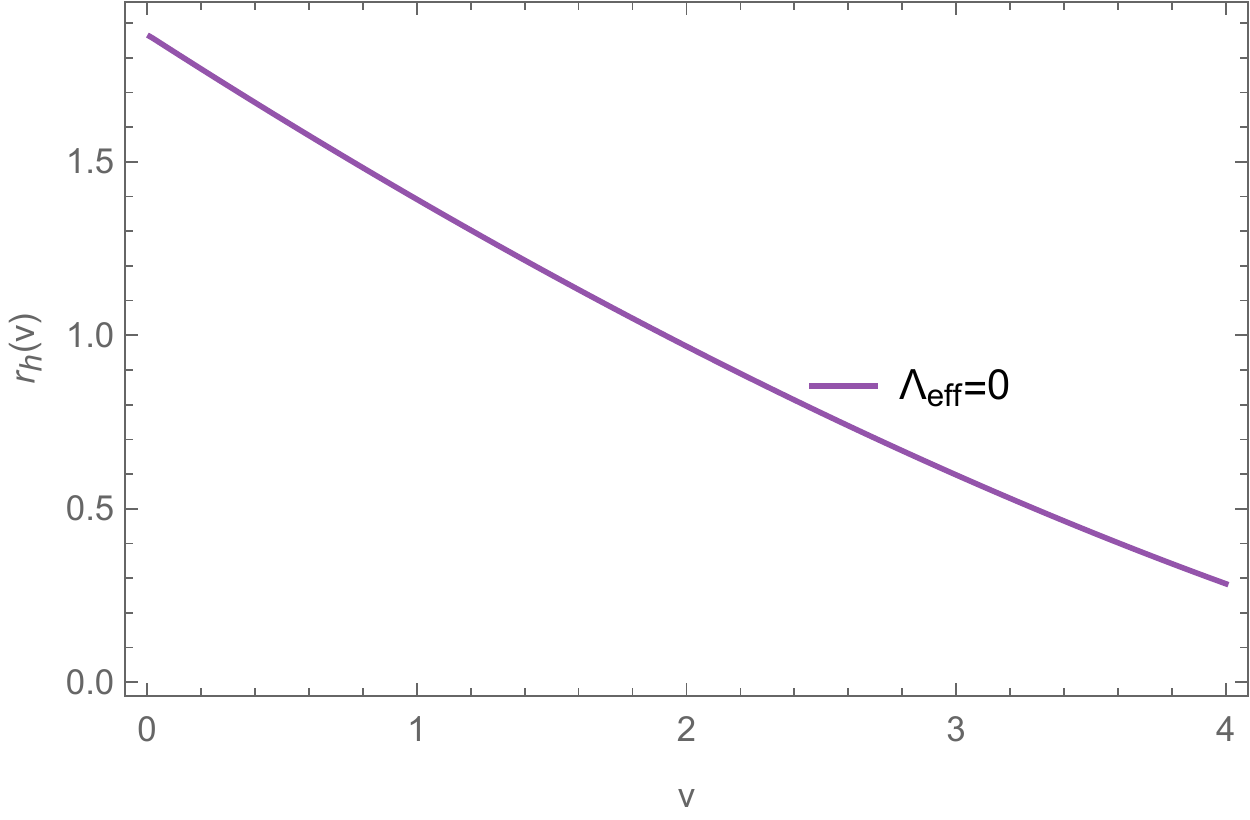}
\caption{The horizon radius $r_h$ as a function of $\nu$ having set $m=1, Q_{m}=0.5 , {\color{blue}\alpha=0.5},$ for asymptotically AdS $\Lambda_{\text{eff}}=-1$ (left) and flat $\Lambda_{\text{eff}}=0$ (right) spacetimes.} \label{eh}
\end{figure}

The system of the field equations (\ref{emt})-(\ref{enmom}) admits an exact magnetically charged solution given by
\begin{eqnarray}
&&\mathcal{A}(r)=0~,\\
&&\phi(r)=\frac{1}{\sqrt{2}}\ln \left(1+\frac{\nu }{r}\right)~,\label{scalarfun}\\
&&b_1(r) =  \sqrt{r(\nu +r)}~,
\end{eqnarray}
while the metric function $b(r)$ is obtained as
\begin{multline}
b(r) =c_1 r (\nu +r)+\frac{\left(2 r-c_2\right) (\nu +2 r)-4 Q_m^2}{\nu ^2}+\frac{8 \alpha  Q_m^4 \left(-\nu ^2+12 r^2+12 \nu  r\right) \left(\nu
   ^2+3 r^2+3 \nu  r\right)}{3 \nu ^6 r^2 (\nu +r)^2}+\frac{2}{\nu ^8} \ln \left(\frac{r}{\nu +r}\right)*\\
 \left(-\nu ^5 r (c_2+\nu ) (\nu
   +r)-2 Q_m^2 r (\nu +r) \left(\nu ^4-24 \alpha  Q_m^2\right) \ln
   \left(\frac{r}{\nu +r}\right)+48 \alpha  \nu  Q_m^4 (\nu +2 r)-2 \nu ^5
   Q_m^2 (\nu +2 r)\right),\label{metric}
\end{multline}
where $c_1,c_2$ are constants of integration and $\nu$ is the scalar charge, also a constant of integration which determines the behavior of the scalar field. For a well behaved scalar field we will impose $\nu>0$. At large distances, the metric function asymptotes to
\begin{equation} b(r\to\infty)\sim 1+\frac{-c_2-\nu }{3 r}+\frac{c_2 \nu +\nu ^2+6
   Q_m^2}{6 r^2}+r^2 \left(c_1+\frac{4}{\nu ^2}\right)+\frac{r \left(c_1 \nu
   ^2+4\right)}{\nu }-\frac{\nu  \left(c_2 \nu +\nu ^2+10
   Q_m^2\right)}{10 r^3}+\mathcal{O}\left(\left(\frac{1}{r}\right)^4\right)~.\end{equation}
We can see that the scalar charge $\nu$ is introducing a new scale in the theory which leads to the appearance of  an effective cosmological constant. Also the generated mass term is given by both an integration constant and the scalar charge $\nu$, hence the scalar field dresses the black hole with a secondary scalar hair.
By redefining the integration constants, the asymptotic relation yields
\begin{equation} b(r\to\infty)\sim 1-\frac{2 m}{r}+\frac{m \nu +Q_m^2}{r^2}-\frac{\Lambda_{\text{eff}} r^2}{3}-\frac{1}{3} r (\Lambda_{\text{eff}} \nu
   )-\frac{\nu  \left(3 m \nu +5
   Q_m^2\right)}{5 r^3}+\mathcal{O}\left(\left(\frac{1}{r}\right)^4\right)~,\end{equation}
where we have set $m=\cfrac{c_2+\nu}{6}$ and $\Lambda_{\text{eff}}=-\Big{(}3c_1+\frac{12}{\nu^2}\Big{)} $. For small $r$ the metric behaves as
\begin{equation}
b(r\to0)\sim -\frac{8 \alpha  Q_m^4}{3 \nu ^4 r^2} + \mathcal{O}(r^{-1})~,
\end{equation}
from which we can deduce that the solution always describes a black hole at least in asymprotically flat or AdS spacetime, due to the fact that $b(r)$ is continuous and changes sign in the range $0<r<\infty$.
In the small scalar hair case ($\nu \to 0$), the metric function yields
\begin{equation} b(r)= \left(1-\frac{2 m}{r}+\frac{Q_m^2}{r^2}-\frac{2 \alpha  Q_m^4}{5
   r^6}-\frac{\Lambda_{\text{eff}} r^2}{3}\right)+\nu
   \left(\frac{m}{r^2}+\frac{6 \alpha  Q_m^4}{5
   r^7}-\frac{Q_m^2}{r^3}-\frac{\Lambda_{\text{eff}} r}{3}\right)+\mathcal{O}\left(\nu
   ^2\right)~.
\end{equation}
Using the metric function (\ref{metric}) and the scalar field function (\ref{scalarfun}) from the system of the field equations (\ref{emt})-(\ref{enmom})
we can specify the scalar potential
\begin{multline}
V(\phi) = \frac{1}{3 \nu ^8}\Bigg(\nu ^8 \Lambda _{\text{eff}} \left(\cosh \left(\sqrt{2} \phi \right)+2\right)-36
   m \nu ^5 \left(\sqrt{2} \phi  \left(\cosh \left(\sqrt{2} \phi \right)+2\right)-3
   \sinh \left(\sqrt{2} \phi \right)\right)-4 \alpha  Q_m^4*\\
\Big(288 \phi ^2+2
   \left(72 \phi ^2+71\right) \cosh \left(\sqrt{2} \phi \right)-432 \sqrt{2} \phi
   \sinh \left(\sqrt{2} \phi \right)+100 \cosh \left(2 \sqrt{2} \phi \right)-14 \cosh
   \left(3 \sqrt{2} \phi \right)+\cosh \left(4 \sqrt{2} \phi \right)-229\Big)\\
+6 \nu
   ^4 Q_m^2 \left(8 \phi ^2+4 \left(\phi ^2+2\right) \cosh \left(\sqrt{2} \phi
   \right)-12 \sqrt{2} \phi  \sinh \left(\sqrt{2} \phi \right)+\cosh \left(2 \sqrt{2}
   \phi \right)-9\right)
\Bigg)~.
\end{multline}
For small $\phi$ we have
\begin{equation} V(\phi) \sim \Lambda _{\text{eff}}+\frac{\phi ^2 \Lambda _{\text{eff}}}{3}+\frac{\phi ^4 \Lambda
   _{\text{eff}}}{18}-\frac{4 \left(\sqrt{2} m\right) \phi ^5}{5 \nu ^3}+\mathcal{O}\left(\phi
   ^6\right)~.\end{equation}
We can also express the potential as a function of $r$
\begin{multline}
V(r) = \frac{1}{6 \nu ^8 r^4 (\nu +r)^4} \\
\bigg(6 \nu ^6 Q_m^2 r^2 (\nu +r)^2 \left(\nu ^2+12 r^2+12 \nu  r\right) +\nu ^6 r^3 (\nu +r)^3 \left(108 m (\nu +2 r)+\Lambda_{\text{eff}} \nu ^2 \left(\nu
   ^2+6 r^2+6 \nu  r\right)\right) \\
-4 \alpha  \nu ^2 Q_m^4 \left(\nu ^6+1332 \nu ^2 r^4+504 \nu ^3 r^3+30 \nu ^4
   r^2+1296 \nu  r^5+432 r^6-6 \nu ^5 r\right) + \\
12 r^3 (\nu +r)^3 \ln \left(\frac{r}{\nu +r}\right) \Big(3 m \nu ^5 \left(\nu ^2+6
   r^2+6 \nu  r\right)+Q_m^2 \left(\nu ^2+6 r^2+6 \nu  r\right) \left(\nu ^4-24
   \alpha  Q_m^2\right) \ln \left(\frac{r}{\nu +r}\right)\\-144 \alpha  \nu
   Q_m^4 (\nu +2 r)+6 \nu ^5 Q_m^2 (\nu +2 r)\Big)
\bigg)~,
\end{multline}
and for small scalar hair ($\nu \to 0$):
\begin{equation} V(r) = \Lambda _{\text{eff}}+\frac{\nu ^2 \left(25 r^8 \Lambda _{\text{eff}}-18 \alpha
   Q_m^4+25 r^4 Q_m^2-30 m r^5\right)}{150 r^{10}}+\mathcal{O}\left(\nu ^3\right)~.\end{equation}
As expected, at zeroth order one obtains the cosmological constant. Its asymptotic behavior at large distances reads
\begin{equation} V(r\to\infty)\sim\Lambda_{\text{eff}}+\frac{\Lambda_{\text{eff}} \nu ^2}{6 r^2} -\frac{\Lambda_{\text{eff}} \nu ^3}{6 r^3}+\mathcal{O}\left(\left(\frac{1}{r}\right)^4\right)~.
\end{equation}
There is also a mass term in the potential
\begin{equation} m^2 =V''(\phi=0)= \frac{2}{3}\Lambda_{\text{eff}}~,\end{equation}
which in the case of AdS spacetime is negative and the scalar field is a tachyon, however in this case it still respects the Breitenlohner-Friedman bound that ensures the perturbative stability of the AdS spacetime \cite{Breitenlohner:1982jf}. The Kretschmann scalar is singular at the origin
\begin{equation}
R_{\mu\nu\chi\psi}R^{\mu\nu\chi\psi}(r\to0) \sim \frac{304 \alpha ^2 Q_m^8}{\nu ^8 r^8}-\frac{7520 \left(\alpha ^2 Q_m^8\right)}{3 \nu ^9 r^7} + \mathcal{O}\left(\frac{1}{r}\right)^6~,
\end{equation}
while it is regular for any $r>0$ and at infinity its behavior is
\begin{eqnarray}
&&R_{\mu\nu\chi\psi}R^{\mu\nu\chi\psi}(r\to\infty) \sim \frac{8 \Lambda_{\text{eff}}^2}{3}+\frac{2 \Lambda_{\text{eff}}^2 \nu ^2}{3 r^2}+\mathcal{O}\left(\left(\frac{1}{r}\right)^3\right)~,\\
&&R_{\mu\nu\chi\psi}R^{\mu\nu\chi\psi}_{\Lambda_{\text{eff}}=0}(r\to\infty) \sim \frac{48 m^2}{r^6}-\frac{8 \left(18 \nu  m^2+\nu ^2 m+12 m Q_m^2\right)}{r^7} + \mathcal{O}\left(\frac{1}{r}\right)^8~.
\end{eqnarray}
Thus the solution is valid for any $r>0$ and describes a black hole in asymptotically (A)dS or flat spacetime for appropriate relations between the parameters. We will focus on the AdS case in order to make comparisons with the uncharged AdS hairy black hole and the flat case which is also of great interest. As can be seen from the definition of $\Lambda_{\text{eff}}$ in order to obtain a flat spacetime the scale introduced by the presence of the scalar field has to be canceled by the integration constant $c_1$.
\begin{figure}[h]
\centering
\includegraphics[width=.40\textwidth]{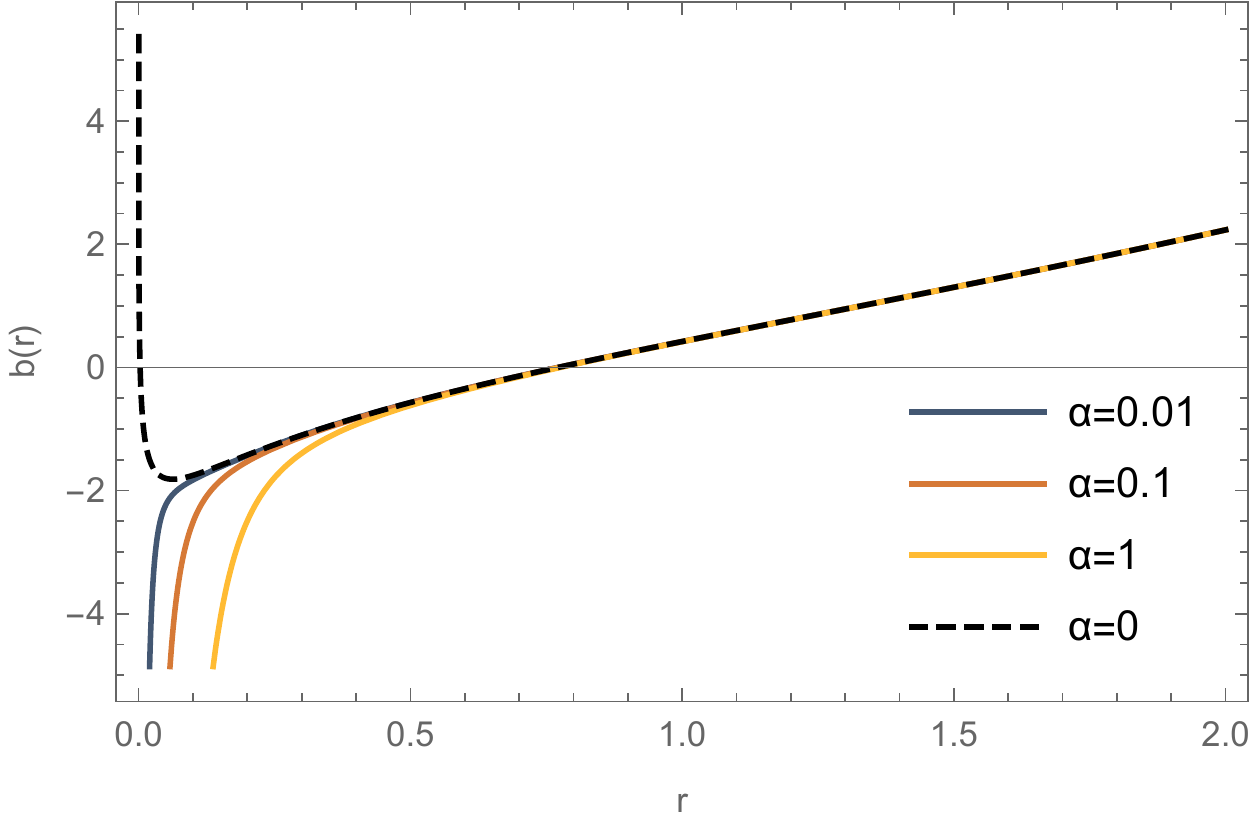}
\includegraphics[width=.40\textwidth]{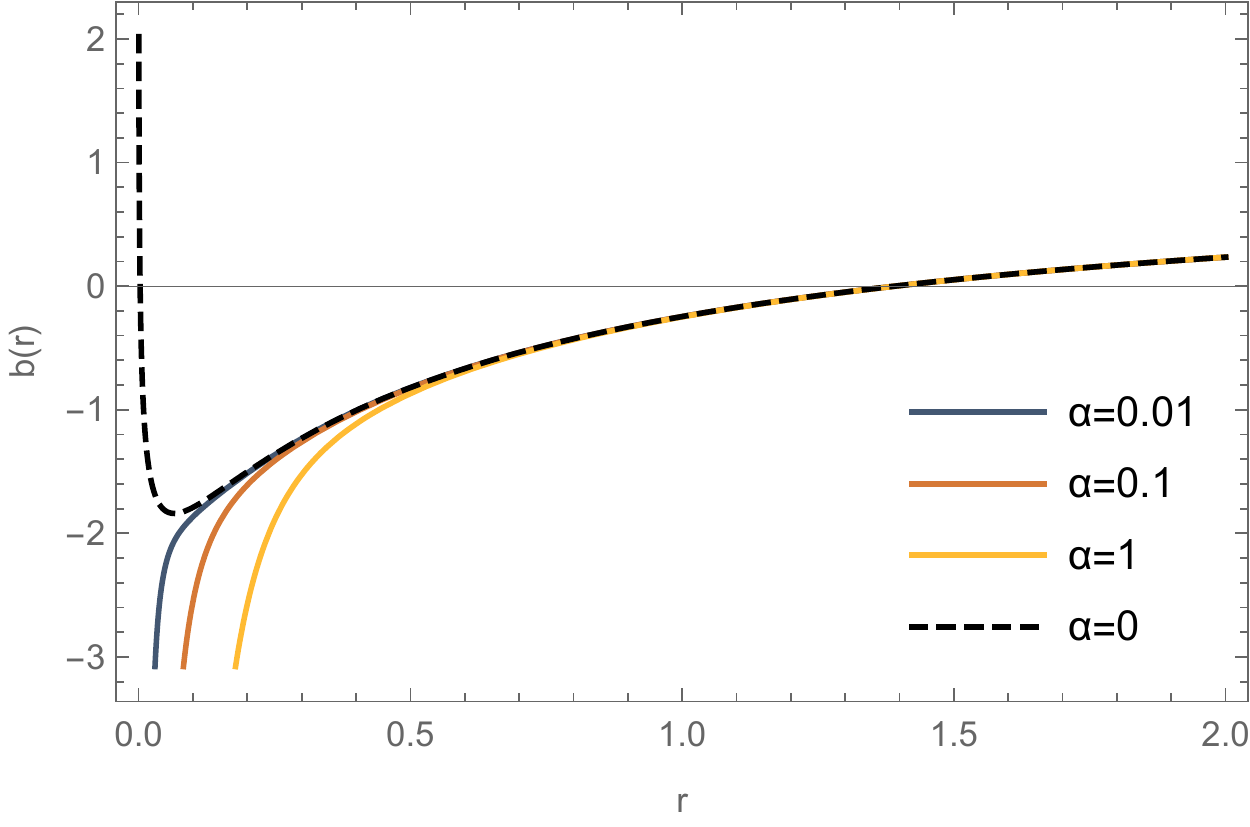}
\includegraphics[width=.40\textwidth]{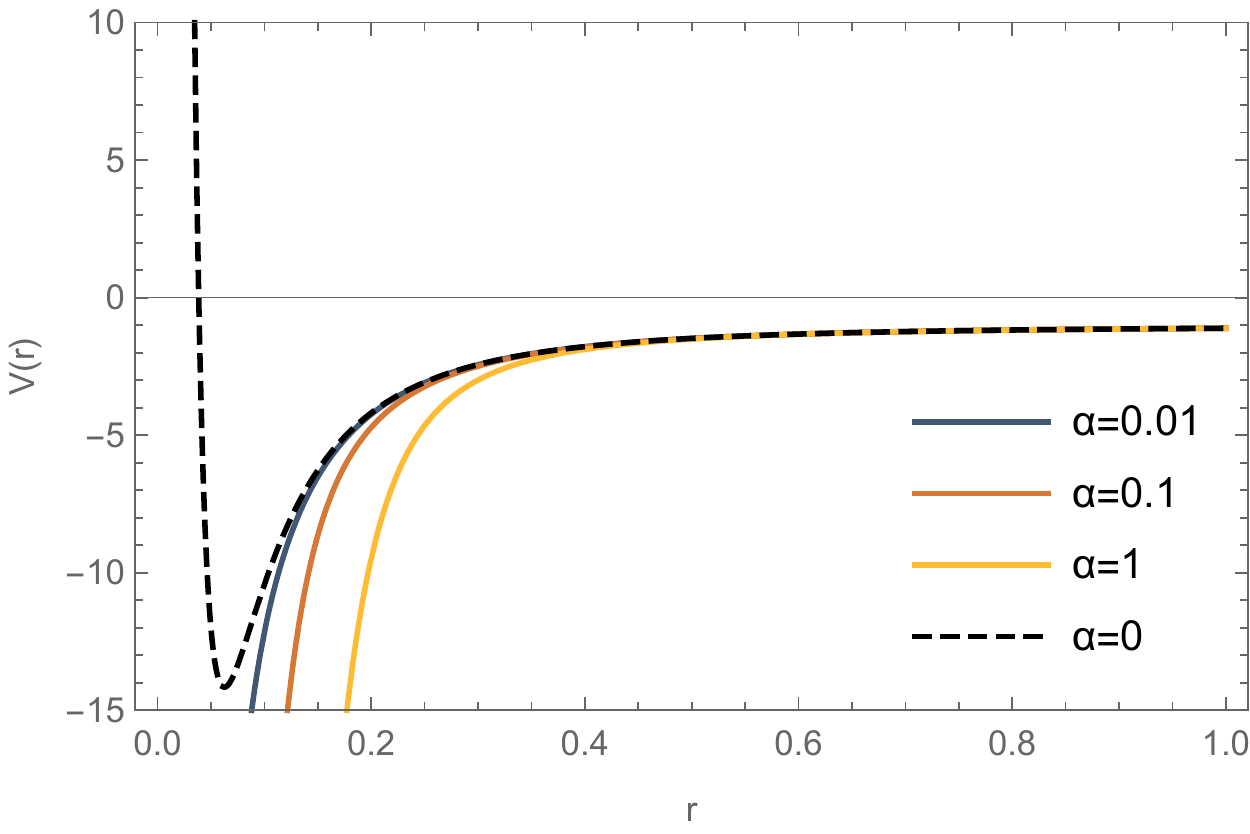}
\includegraphics[width=.40\textwidth]{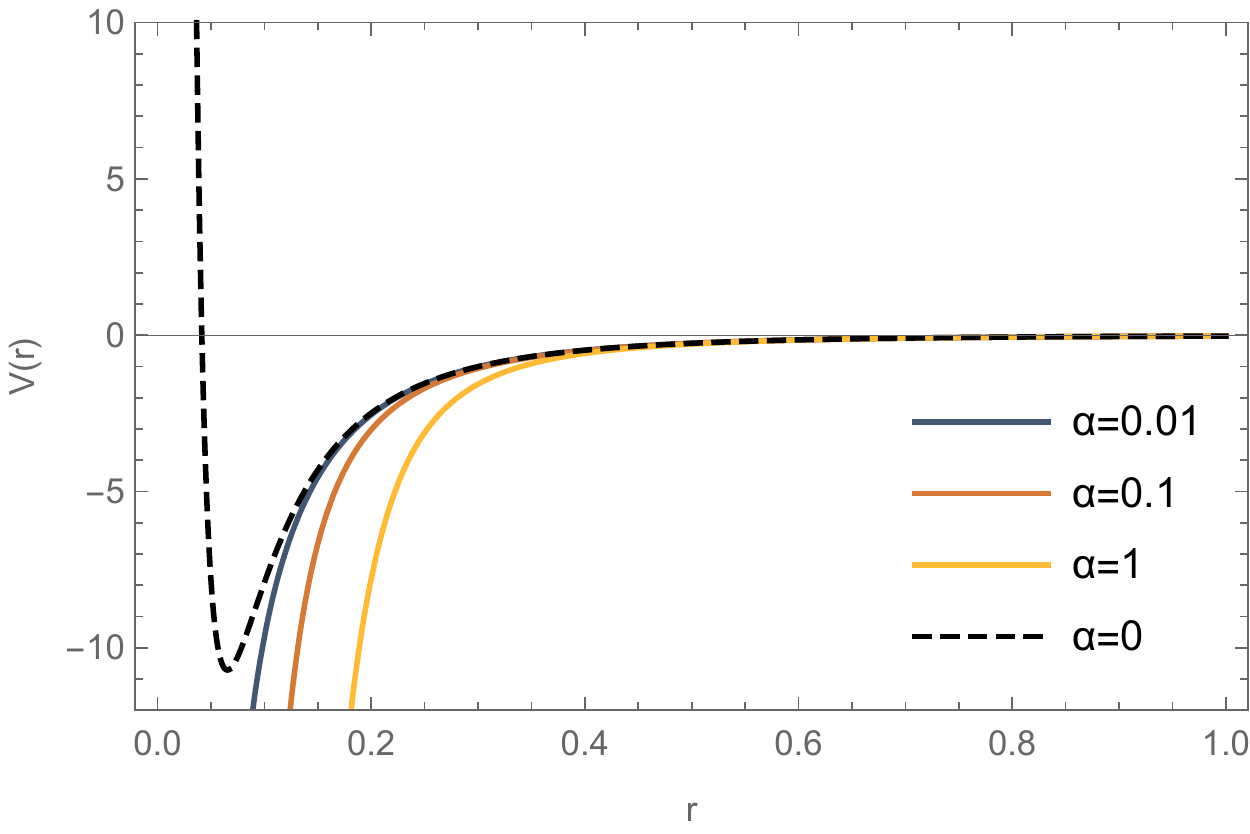}
\caption{Hairy black hole configurations for asymptotically AdS $\Lambda_{\text{eff}}=-1$ (left) and flat $\Lambda_{\text{eff}}=0$ (right) spacetimes, where we have fixed $m=1, Q_{m}=0.5 , \nu=1,$ while changing the Euler-Heisenberg parameter $\alpha$.} \label{alpha}
\end{figure}
\begin{figure}[h]
\centering
\includegraphics[width=.40\textwidth]{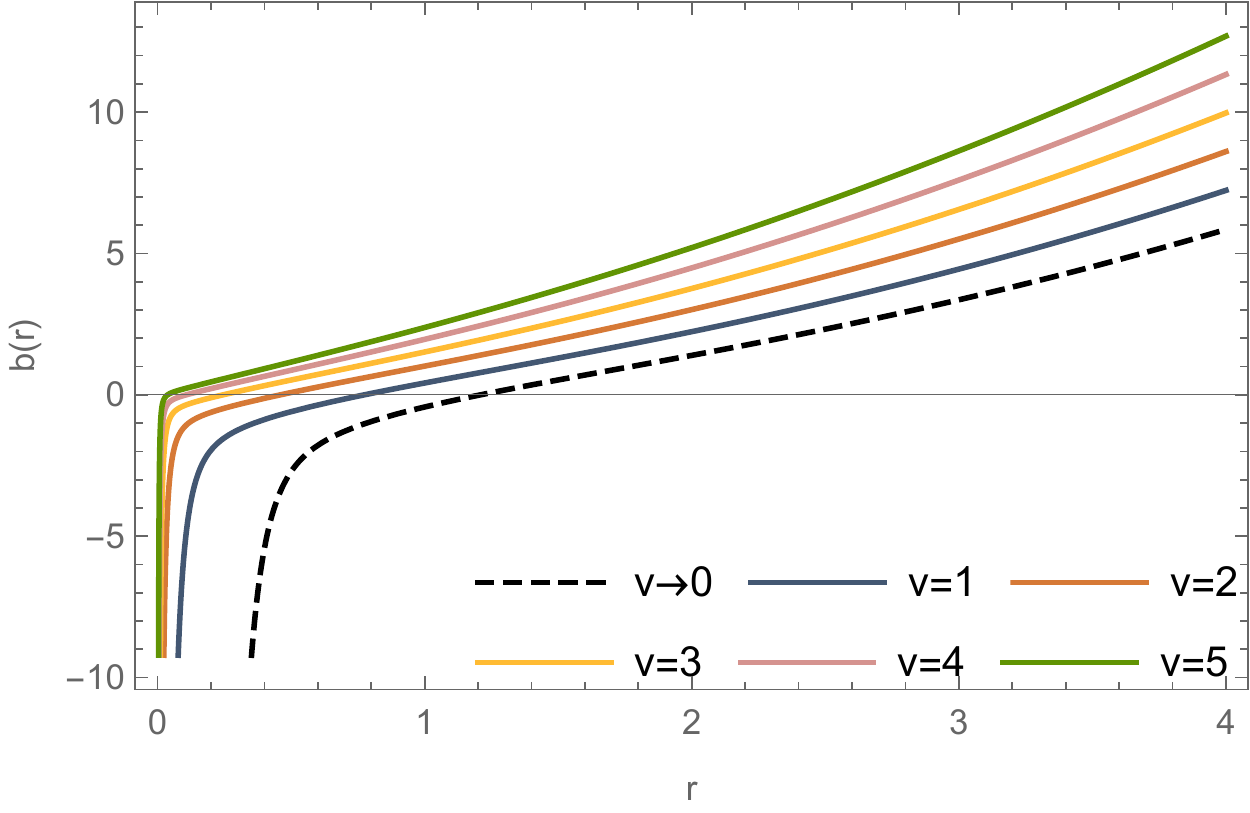}
\includegraphics[width=.40\textwidth]{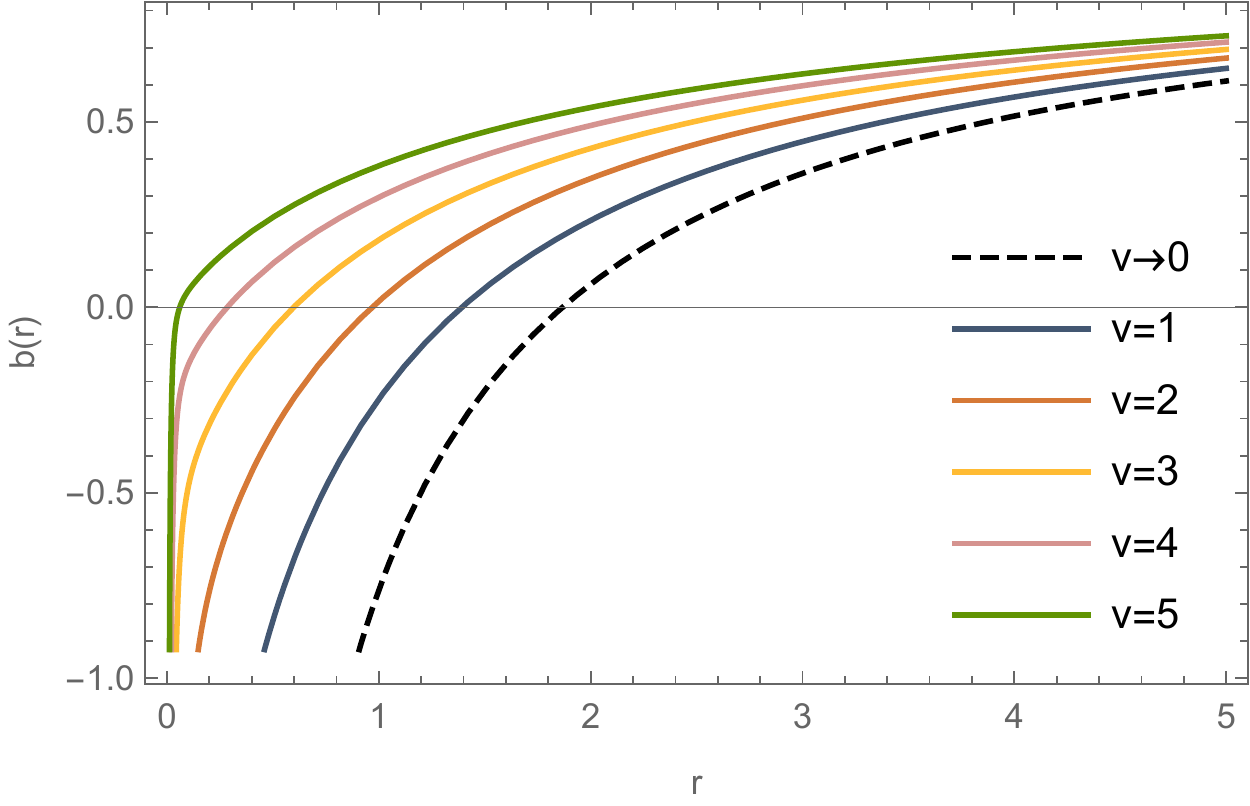}
\includegraphics[width=.40\textwidth]{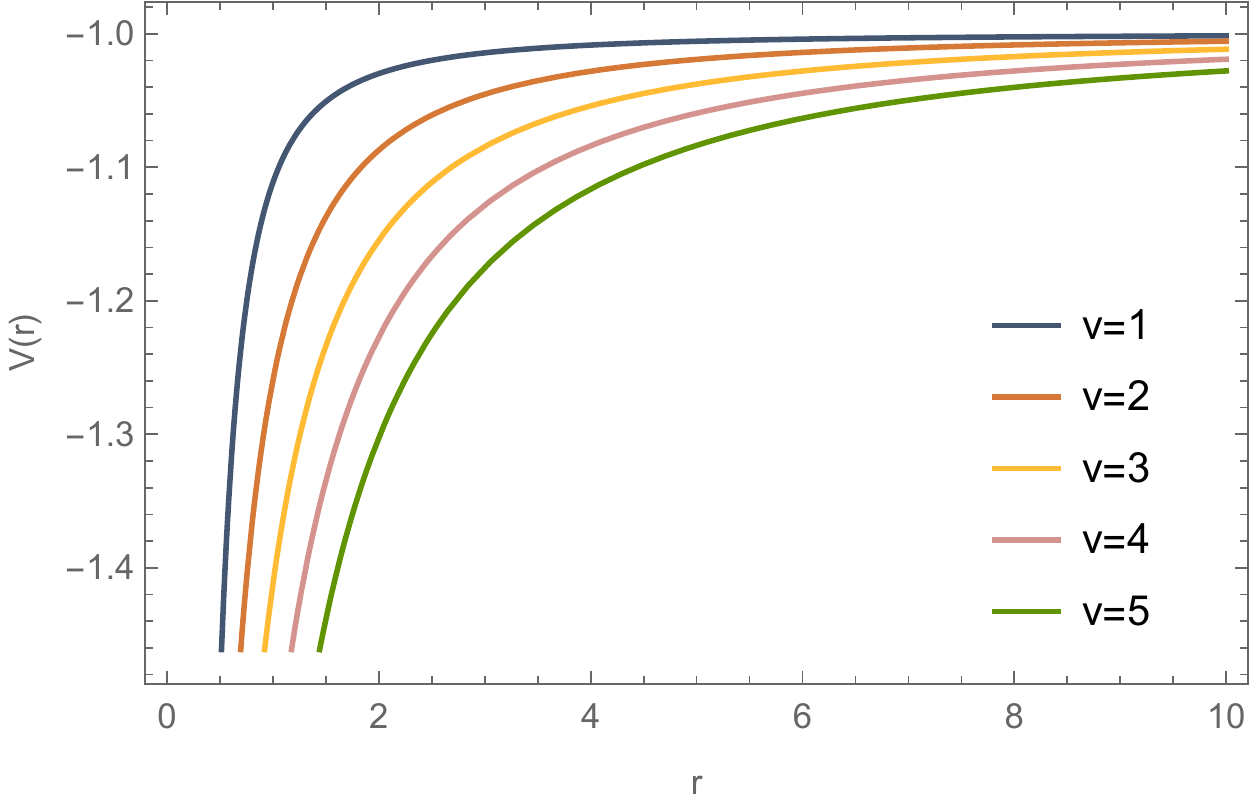}
\includegraphics[width=.40\textwidth]{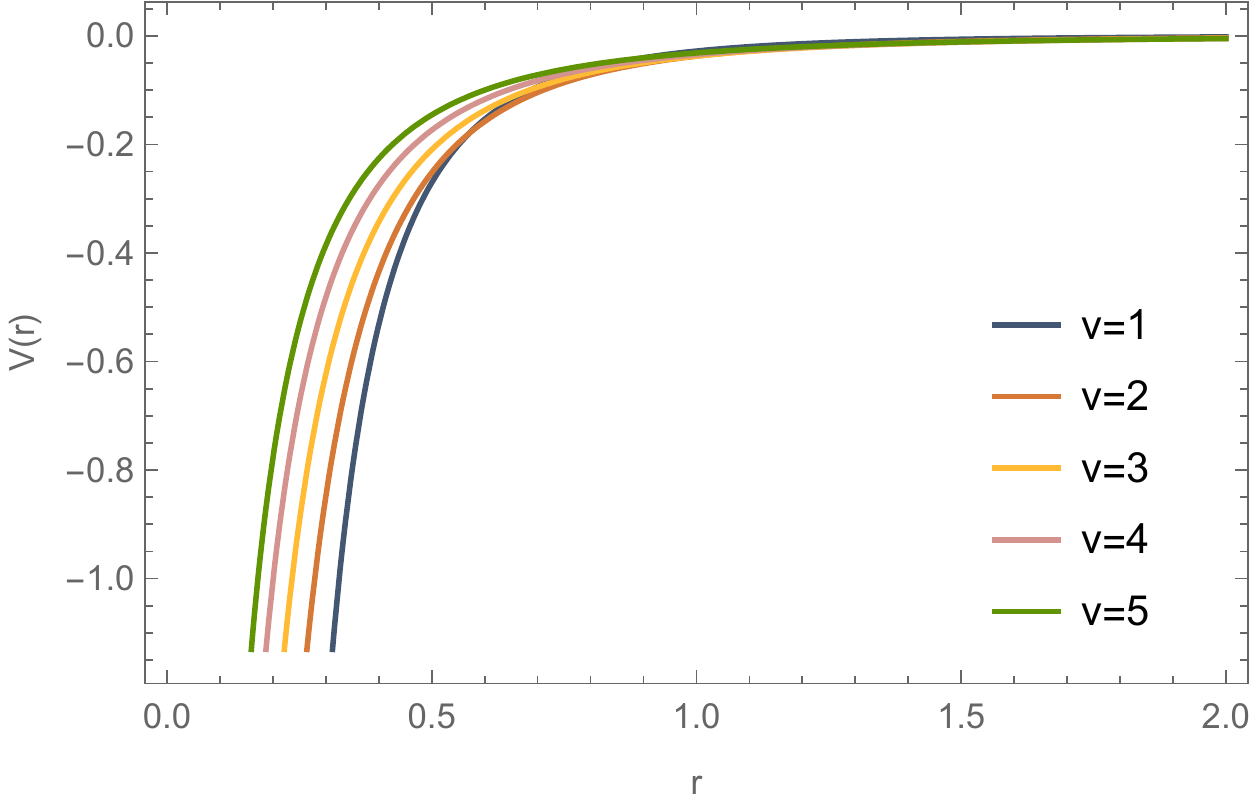}
\caption{Hairy black hole configurations for asymptotically AdS $\Lambda_{\text{eff}}=-1$ (left) and flat $\Lambda_{\text{eff}}=0$ (right) space-times, where we have fixed $m=1, Q_{m}=0.5 , {\color{blue}\alpha=0.5},$ while changing the scalar charge $\nu$.} \label{nu}
\end{figure}
In Fig. \Ref{alpha} we plot the metric function $b(r)$ and the potential $V(r)$ for the asymptotically AdS and flat spacetimes for a fixed scalar charge while changing $\alpha$. The $\alpha=0$ case differs in structure with the $\alpha \neq 0$ cases, having an inner and an event horizon. The Euler-Heisenberg parameter $\alpha$ does not affect the horizon radius of the black hole as we can see. Moreover, the potentials are negative in order to support the hairy structure and violate the no-hair theorem. It is worth noting that, as we can see from the figures, $\alpha$ acts in favour of the no-hair theorem, since the existence of $\alpha$ ensures a negative potential everywhere, while for $\alpha=0$ there is a small region where the potential can be positive. In Fig. \Ref{nu} we also plot $b(r),V(r)$ while chancing $\nu$ having set $\alpha=0.5$. We can see that for bigger $\nu$ (stronger scalar field) the black hole horizon radius is smaller. We also evaluate numerically and plot in Fig. \ref{eh} the horizon radius as a function of $\nu$ for both AdS and flat cases to visualise how the horizon changes as a function of the scalar charge.

\begin{figure}[h]
\includegraphics[width=8cm]{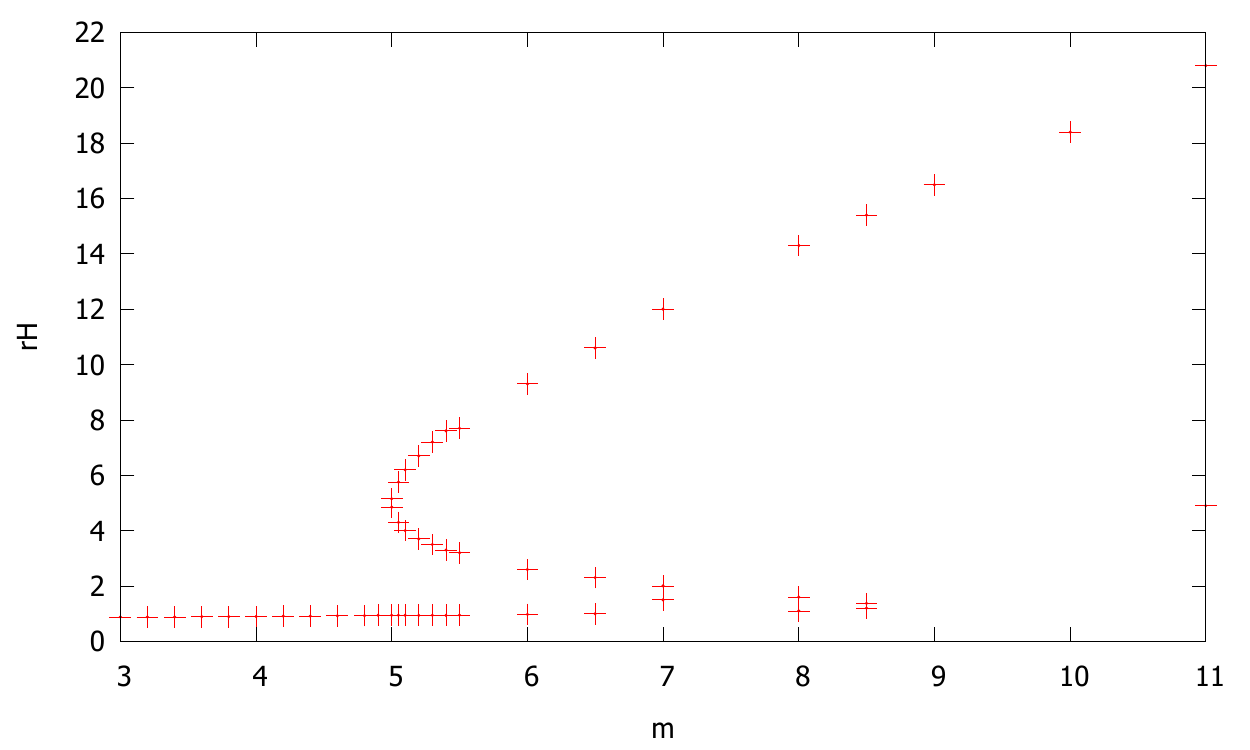}
\includegraphics[width=8cm]{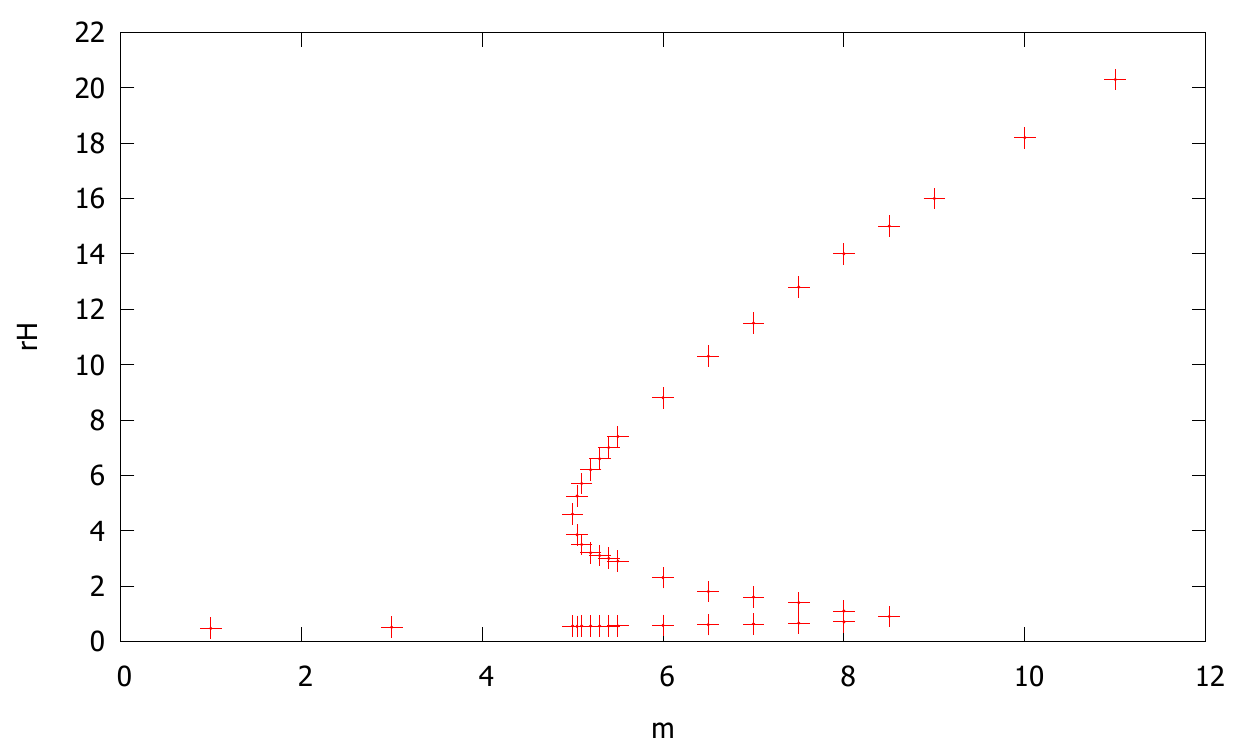}
\centering
\caption{Horizons versus Black hole mass. The asymptotically flat case $(\Lambda_{\text{eff}}=0)$ is depicted. The Euler-Heisenberg parameter $\alpha$ equals $0.05.$ Left panel: No scalar field $(\nu=0).$ Right panel: Small scalar field $(\nu=1).$}
\centering
\label{qm=5_Lambda=0_a=005}
\end{figure}

To have a better  understanding  of the hairy Euler-Heisenberg black  hole we found, we studied the horizon structure  of the various solutions and their dependence on  the gravitational mass. In Fig. \ref{qm=5_Lambda=0_a=005}  we show the dependence of the horizons on the gravitational mass for $\Lambda_{\text{eff}}=0.$   The Euler-Heisenberg parameter is small $(\alpha=0.05)$ and the magnetic field is fairly large, $Q_m=5.$ We find that, in the interval of black hole masses between 5.5 and 8.0 (or 8.5) there are three possible solutions of the metric function, signalling one outer  the two inner horizons. The influence of the scalar field is not large. The extreme solutions, where we get just two horizons correspond to vanishing temperatures.

\begin{figure}[h]
\includegraphics[width=8cm]{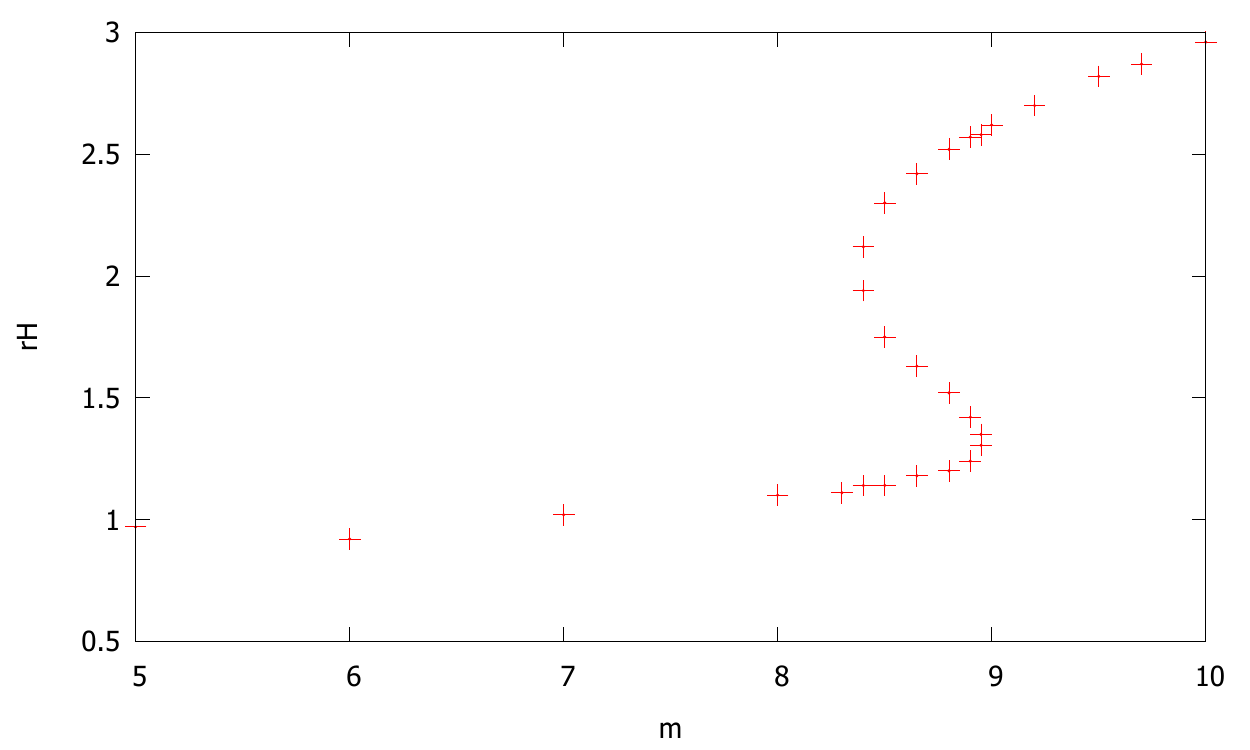}
\includegraphics[width=8cm]{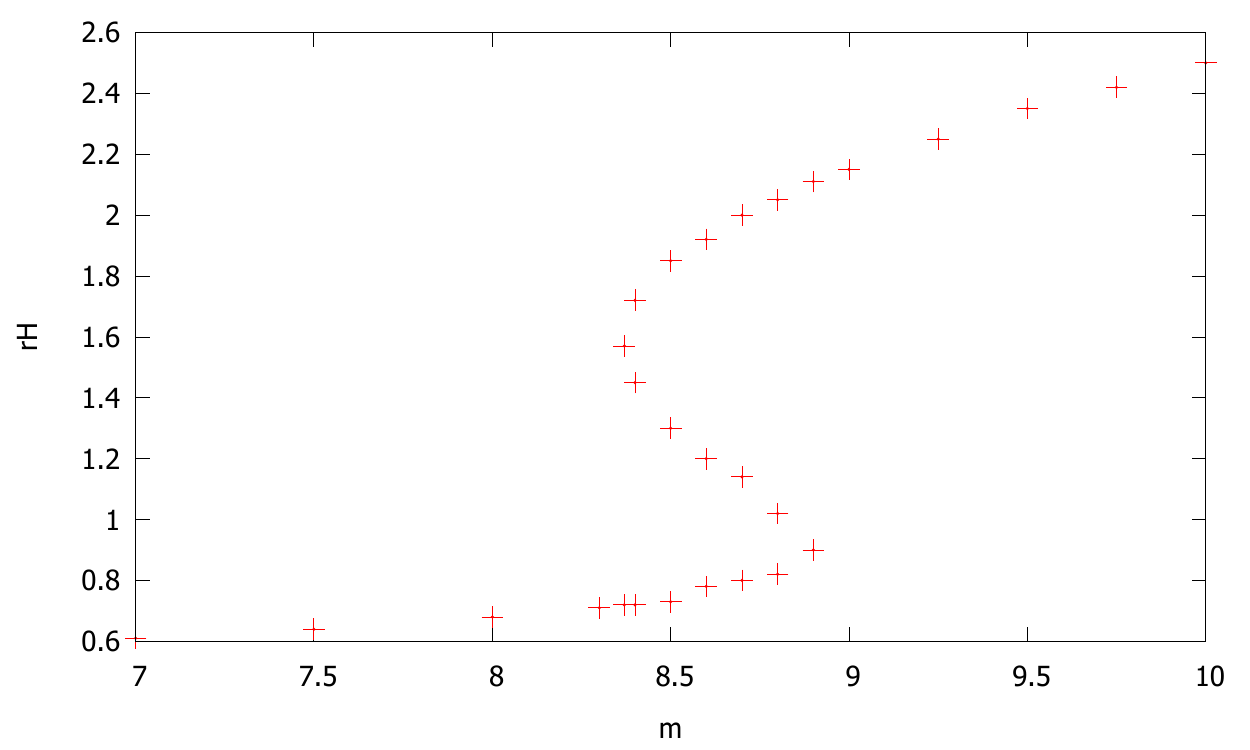}
\centering
\caption{Horizons versus Black hole mass. The asymptotically AdS case $(\Lambda_{\text{eff}}=-1)$ is depicted. The Euler-Heisenberg parameter $\alpha$ equals $0.05.$ Left panel: No scalar field $(\nu=0).$ Right panel: Small scalar field $(\nu=1).$}
\centering
\label{qm=5_Lambda=-1_a=005}
\end{figure}

In Fig. \ref{qm=5_Lambda=-1_a=005} we show the horizon structures for asymptotically AdS spaces with $(\Lambda_{\text{eff}}=-1)$. The Euler-Heisenberg parameter is small $(\alpha=0.05)$ and the magnetic field is fairly large, $Q_m=5.$ We find that, in the interval of black hole masses between 8.5 and 9.0, there are three possible roots again. The influence of the scalar field is not large here either. We observe that the range of variation of the horizons is considerably smaller than the previous case: it varies between 1.0 and 3.5, which is an order of magnitude smaller than before. The extreme solutions, where we get just two solutions correspond to vanishing temperatures.

\begin{figure}[h]
\includegraphics[width=8cm]{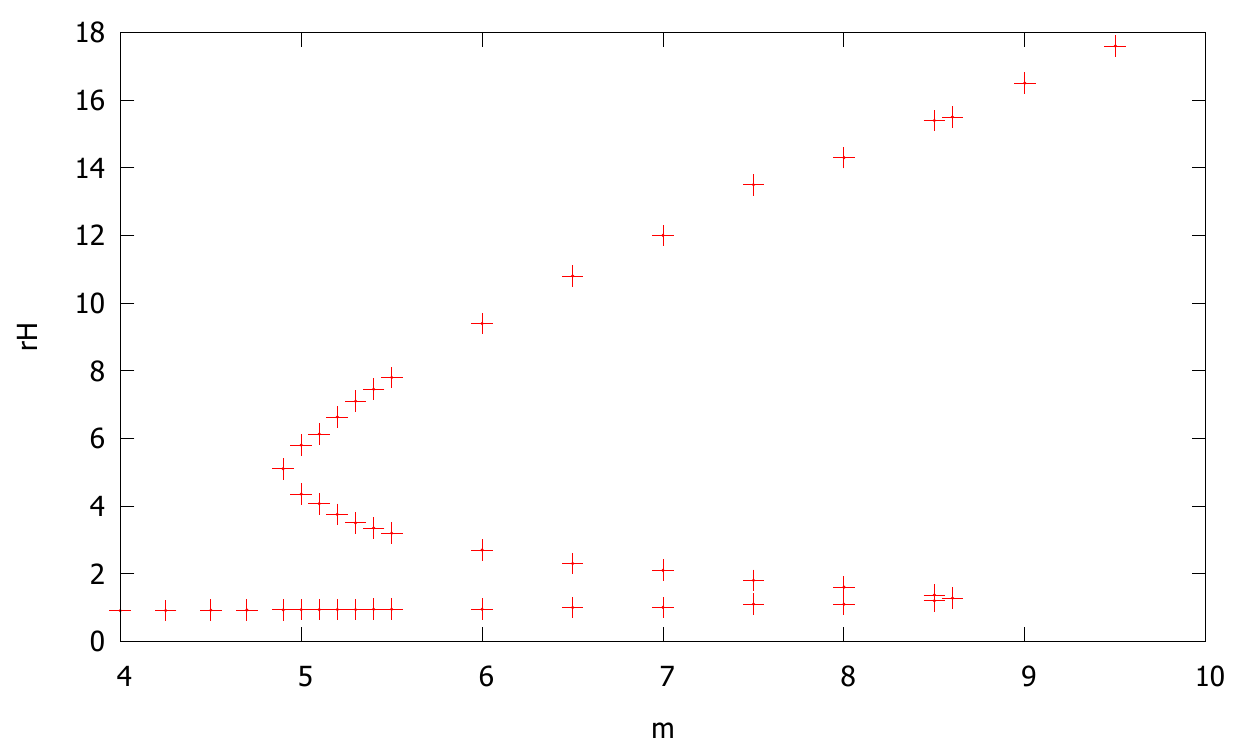}
\includegraphics[width=8cm]{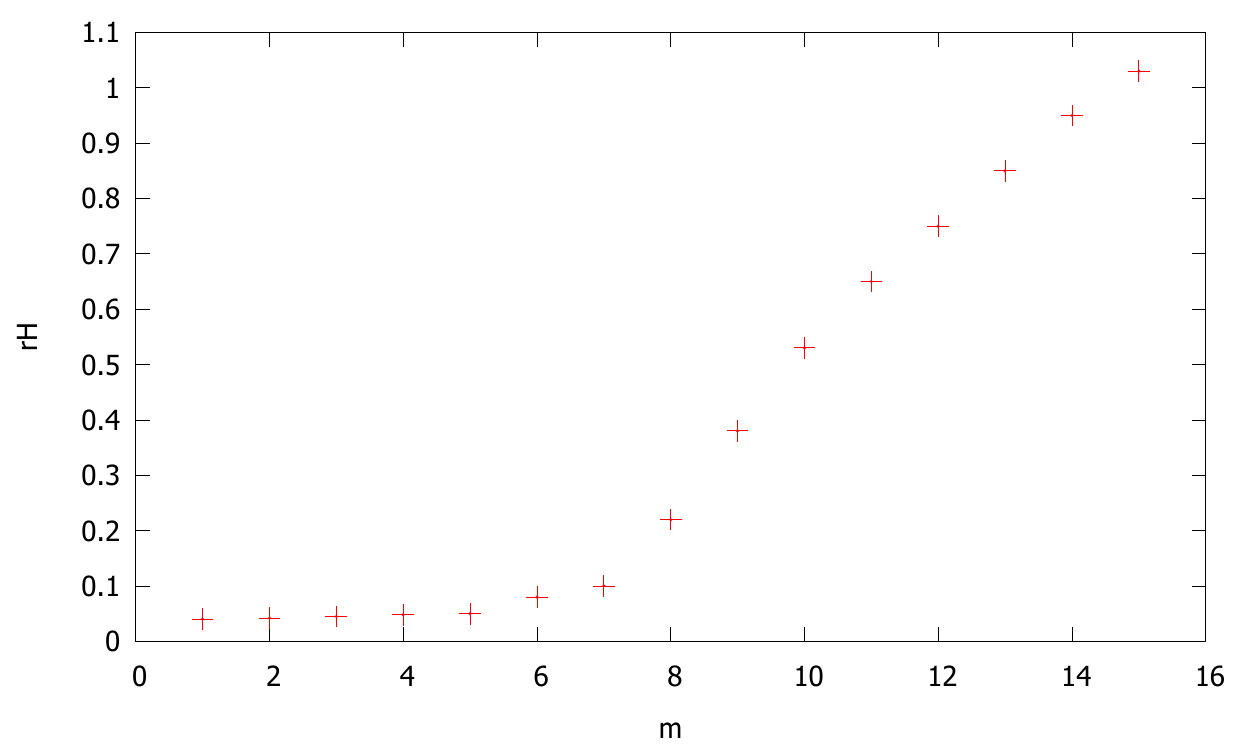}
\centering
\caption{Horizons versus Black Hole mass. Large scalar field charge $(\nu=10).$ Left panel: Asymptotically flat case $(\Lambda_{\text{eff}}=0).$ Right panel: Asymptotically AdS case $(\Lambda_{\text{eff}}=-1).$}
\centering
\label{qm=5_nu=10_a=005}
\end{figure}

Finally in Fig. \ref{qm=5_nu=10_a=005} we show once more the behaviour of the horizons for a large scalar charge $(\nu=10).$ It contains both the asymptotically flat case and the asymptotically AdS case. We see the remarkable characteristic that the horizon (just one solution for each $m$) starts off with very small values, while, when $m$ is large enough, it jumps to a much larger value. That is a large black hole get suddenly large horizon values. In addition, the horizon radii for $\Lambda_{\text{eff}}=-1$ are almost one order of magnitude smaller than the ones for $\Lambda_{\text{eff}}=0.$

To summarise our results: The structure of three roots, as well as the existence of points with $T=0,$ appears when $a$ is small. To have this behaviour of the horizons, the magnetic charge $Q_m$ should be large enough. The structure disappears when $\Lambda_{\text{eff}}$ takes on large negative values. The horizon radius is getting large when the scalar charge and the gravitational mass are large.

The  no-hair theorem by Bekenstein, states that for an asymptotically flat spacetime, a positive definite potential cannot violate the no-hair theorem. For this reason we multiply the Klein-Gordon equation (\ref{KG}) by $V(\phi)$ and we integrate over the black hole exterior region
\begin{equation} \int d^4x \sqrt{-g}  \Big(V(\phi)\Box \phi - V(\phi)V'(\phi)\Big) = 0 \to \int d^4x \sqrt{-g} \Big(\nabla_{\mu}\left(V(\phi)\nabla^{\mu}\phi\right) - V'(\phi) \nabla_{\mu}\phi\nabla^{\mu}\phi -V(\phi) V'(\phi)\Big)=0~.\end{equation}
For an asymptotically flat spacetime, we can ignore the first term which is a total derivative and this relation becomes
\begin{equation} \int d^4x \sqrt{-g}V'(\phi)\Big(\nabla_{\mu}\phi\nabla^{\mu}\phi +V(\phi) \Big)=0~.\label{nohair} \end{equation}
It is clear that the kinetic term above is always positive outside the black hole region. In order for the integral to be zero, we want a negative potential in order to counterbalance the positive kinetic term, which will result in a zero area between the curve of the integrand and the $r$ axis.
The presence of the scalar field  introduces a matter distribution outside the horizon of the black hole. The condition (\ref{nohair})  guarantees that
the kinetic energy of the scalar field has to counterbalance the potential energy  of  the scalar field in order to have a stable  matter distribution  outside the horizon of the black hole. Therefore we have to find regions of spacetime where the potentials are negative to violate the no-hair theorem and to support the hairy structure.

\section{Special Cases for Black Hole Solutions}
\label{sec2}

In this section we will present special cases for the black hole solutions we found in the previous section depending on the choice of the parameters.

For the case $\nu\to0$ we have the Euler-Heisenberg black hole \cite{Yajima:2000kw}
\begin{equation}
b(r) =1-\frac{2 m}{r}-\frac{2 \alpha  Q_m^4}{5 r^6}+\frac{Q_m^2}{r^2}-\frac{\Lambda_{\text{eff}} r^2}{3}~,
\end{equation}
while the potential gives the cosmological constant $V=\Lambda_{\text{eff}}.$ As it is expected, because the scalar field is decoupled, if we set the Euler-Heisenberg parameter equal to zero we can obtain the (A)dS RN spacetime, the magnetically charged RN spacetime by also setting $\Lambda_{\text{eff}}=\alpha=0$, and the Schwarzschild one by further imposing $Q_m=0$.

For the case  $\alpha=0$ we obtain novel magnetically charged hairy black hole solutions where the metric function is given by
\begin{multline}
b(r) = 1-\frac{4 Q_m^2}{\nu ^2}-\frac{6 m (\nu +2 r)}{\nu ^2}-\frac{1}{3} \Lambda_{\text{eff}} r (\nu +r) \\
-\frac{4}{\nu ^4} \ln \left(\frac{r}{\nu +r}\right) \left(3 m \nu  r (\nu +r)+\nu  Q_m^2 (\nu +2 r)+Q_m^2 r (\nu +r) \ln \left(\frac{r}{\nu +r}\right)\right)~,
\end{multline}
while the potential will be given by
\begin{multline}
V(\phi) = \frac{1}{3 \nu ^8}\Bigg(\nu ^8 \Lambda _{\text{eff}} \left(\cosh \left(\sqrt{2} \phi \right)+2\right)-36
   m \nu ^5 \left(\sqrt{2} \phi  \left(\cosh \left(\sqrt{2} \phi \right)+2\right)-3
   \sinh \left(\sqrt{2} \phi \right)\right) \\
+6 \nu ^4 Q_m^2 \left(8 \phi ^2+4
   \left(\phi ^2+2\right) \cosh \left(\sqrt{2} \phi \right)-12 \sqrt{2} \phi  \sinh
   \left(\sqrt{2} \phi \right)+\cosh \left(2 \sqrt{2} \phi \right)-9\right)\Bigg)~.\end{multline}

For the case $\alpha=Q_m=0$  we turn back to the well known asymptotically AdS black hole solutions with a scalar hair \cite{Gonzalez:2013aca}
where the metric function will be given by
\begin{equation}
b(r) = 1-\frac{1}{3} r \Lambda _{\text{eff}} (\nu +r)-\frac{6 m (\nu +2 r)}{\nu ^2}-\frac{12 m
   r}{\nu ^3} (\nu +r) \ln \left(\frac{r}{\nu +r}\right)~,
\end{equation}
with potential
\begin{equation}
V(\phi) = \frac{1}{3} \Lambda _{\text{eff}} \left(\cosh \left(\sqrt{2} \phi
   \right)+2\right)-\frac{12 m \left(\sqrt{2} \phi  \left(\cosh \left(\sqrt{2} \phi
   \right)+2\right)-3 \sinh \left(\sqrt{2} \phi \right)\right)}{\nu ^3}~.
\end{equation}

\section{Thermodynamics} \label{sec3}

\begin{figure}[h]
\centering
\includegraphics[width=.40\textwidth]{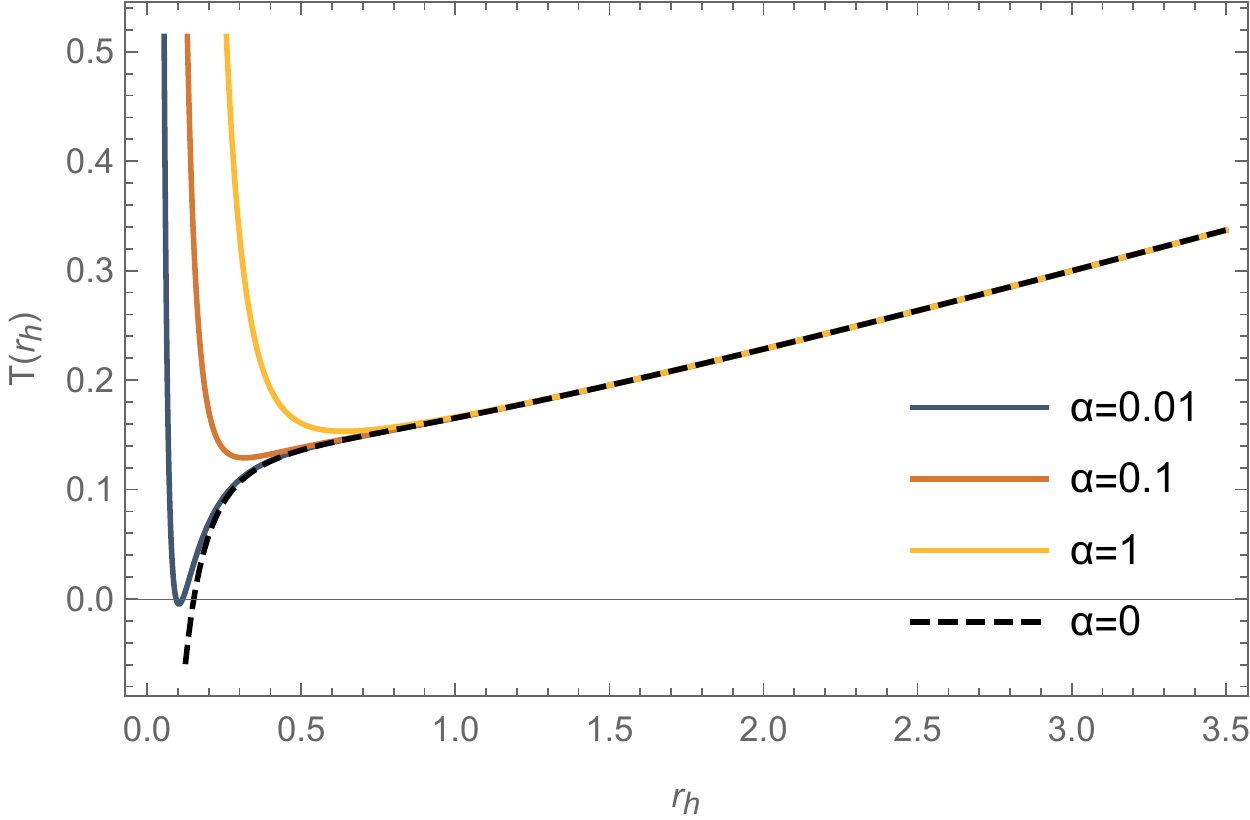}
\includegraphics[width=.40\textwidth]{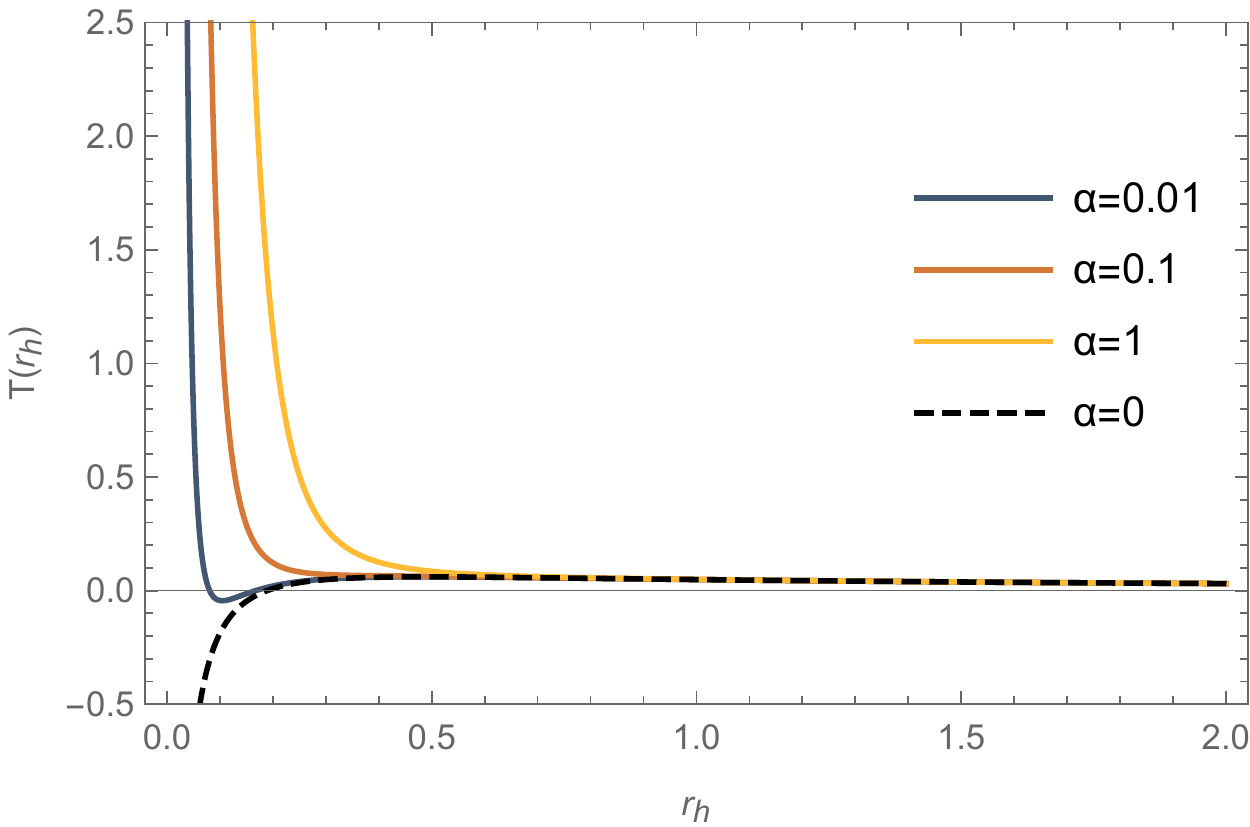}
\caption{The temperature of the hairy black hole configurations for asymptotically AdS $\Lambda_{\text{eff}}=-1$ (left) and flat $\Lambda_{\text{eff}}=0$ (right) spacetimes, where we have fixed $Q_{m}=0.5 , \nu=1,$ while changing the Euler-Heisenberg parameter $\alpha$.} \label{temp}
\end{figure}

In this Section we will discuss the thermodynamical properties of the hairy  black hole solution. We will study the  temperature first.  To do so we perform a Wick rotation $t\to -i\tau,$ and move to Euclidean time. Imposing periodicity of the Euclidean time we can obtain the black hole temperature as
\begin{multline} T(r_h) =\frac{b'(r_h)}{4\pi}= -\frac{1}{12 \pi  \nu ^5 r_h^3 (\nu +r)^3 \left(\nu  (\nu +2 r_h)+2 r_h (\nu +r_h) \ln
   \left(\frac{r_h}{\nu +r_h}\right)\right)}\\
\Bigg(-8 \alpha  \nu ^2 Q_m^4 \left(\nu ^2-6 r_h^2-3 \nu  r_h\right) \left(2 \nu ^2+6
   r_h^2+9 \nu  r_h\right)+12 \nu ^6 Q_m^2 r_h^2 (\nu +r_h)^2+\nu ^6 r_h^3
   \left(\Lambda_{\text{eff}} \nu ^2+12\right) (\nu +r_h)^3\\
+6 r_h (\nu +r_h) \ln \left(\frac{r_h}{\nu +r_h}\right) \bigg(\nu  (\nu +2 r_h) \left(8 \alpha
   Q_m^4 \left(-\nu ^2+6 r_h^2+6 \nu  r_h\right)+\nu ^4 r_h^2 (\nu +r_h)^2\right)\\-2
   Q_m^2 r_h^2 (\nu +r_h)^2 \left(\nu ^4-24 \alpha  Q_m^2\right) \ln
   \left(\frac{r_h}{\nu +r_h}\right)\bigg)\Bigg)~,
\end{multline}
where we have already substituted the mass parameter using the horizon condition $b(r_h)=0$ and $r_h$ denotes the event horizon. For small black holes, the Euler-Heisenberg parameter $\alpha$ plays a decisive role since
\begin{equation} T(r_h\ll1) \sim \frac{4 \alpha  Q_m^4}{3 \pi  \nu ^4 r_h^3} + \mathcal{O}\left(\frac{\ln(r_h)}{r_h^2}\right)~,\end{equation}
while for large black holes the effect of $\alpha$ is negligible
\begin{equation} T(r_h \gg 1) \sim -\frac{\Lambda_{\text{eff}} r_h}{4
   \pi } -\frac{\Lambda_{\text{eff}} \nu }{8 \pi }+ \frac{\Lambda_{\text{eff}} \nu ^2+20}{80 \pi  r_h} + \mathcal{O}\left(\frac{1}{r_h^2}\right)~.
\end{equation}
We can see this behaviour in Fig. \ref{temp} where we plot the temperature of the black hole while changing the Euler-Heisenberg parameter $\alpha$. We observe that $\alpha$ increases the temperature in the case of small black holes.
\begin{figure}[H]
\centering
\includegraphics[width=.40\textwidth]{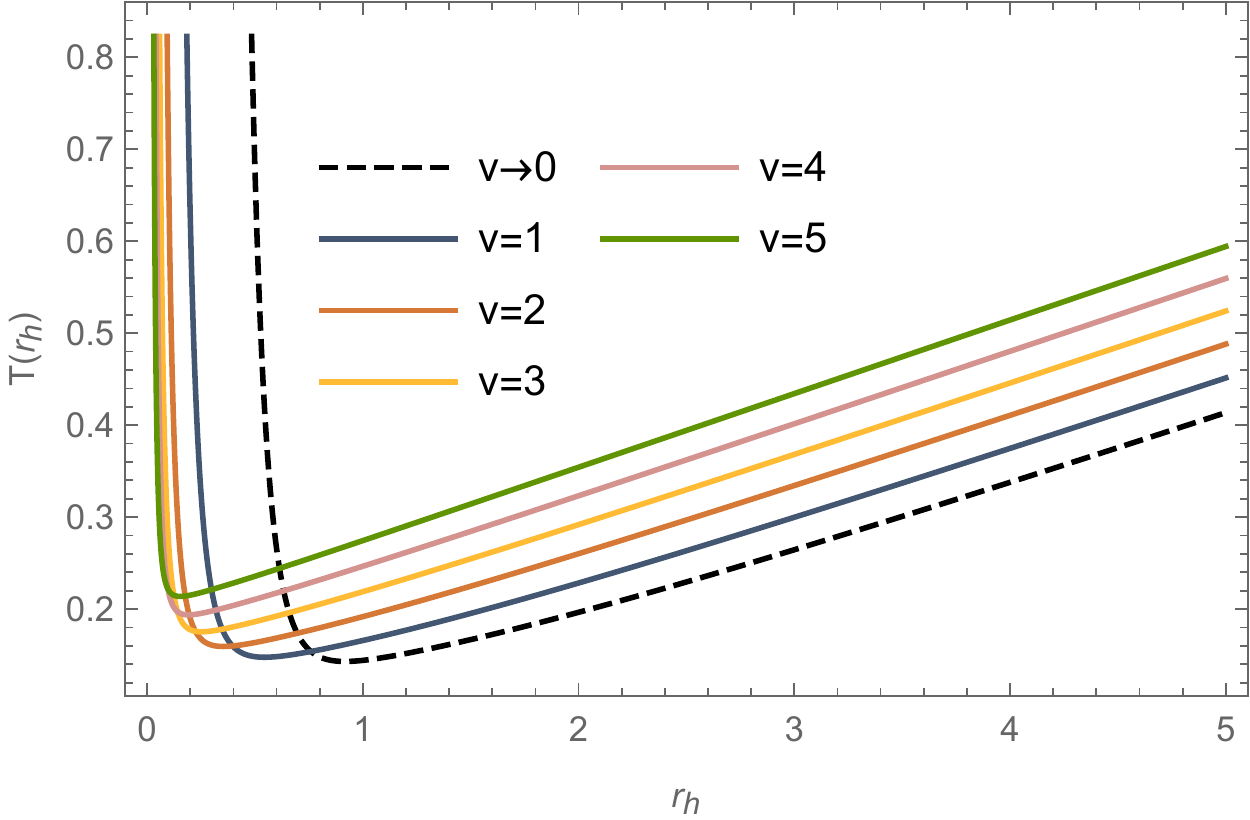}
\includegraphics[width=.40\textwidth]{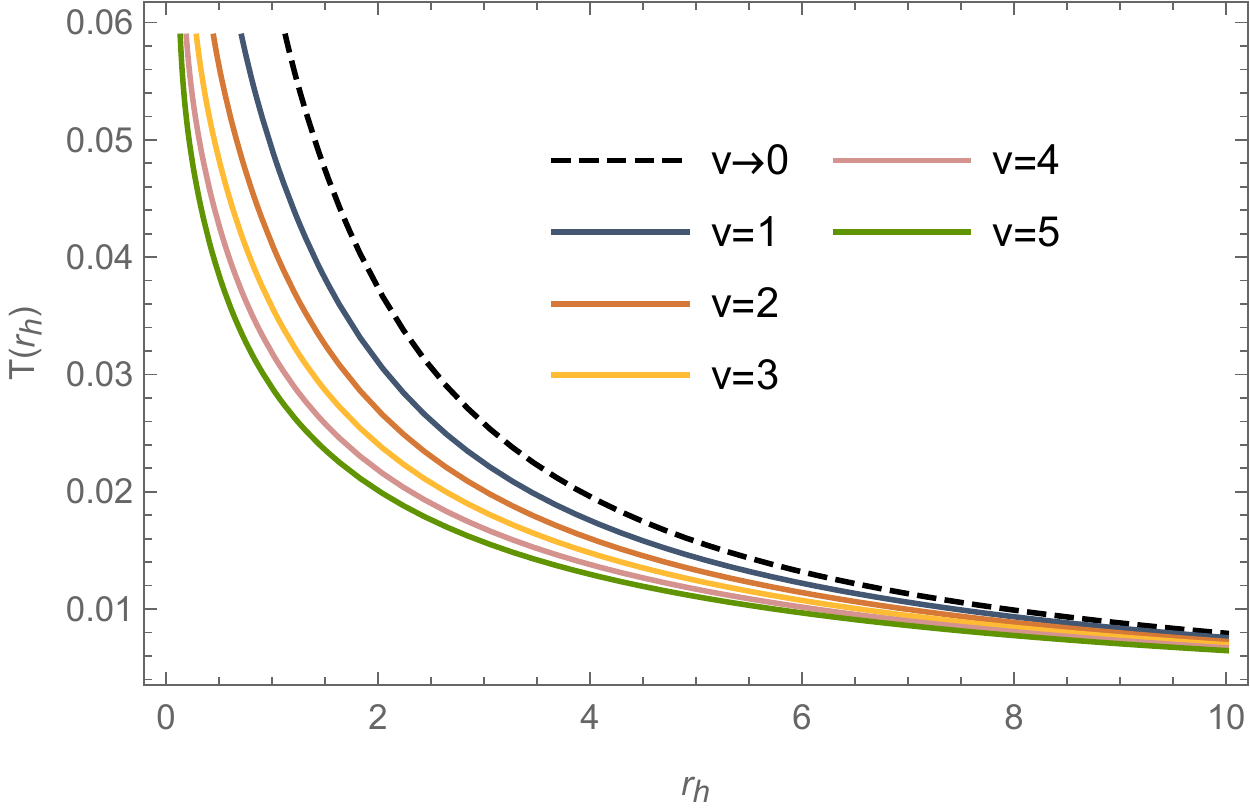}
\caption{The temperature of the hairy black hole configurations for asymptotically AdS $\Lambda_{\text{eff}}=-1$ (left) and flat $\Lambda_{\text{eff}}=0$ (right) spacetimes, where we have fixed $Q_{m}=0.5 , \alpha=0.5,$ while changing the scalar charge $\nu$.} \label{temp1}
\end{figure}
In Fig. \ref{temp1} we plot the black hole temperature having fixed the Euler-Heisenberg parameter $\alpha=0.5$, while we vary the scalar charge of the solution both for asymptotically AdS and flat cases. The temperature of the asymptotically AdS case develops a minimum which can be obtained numerically. For example for $Q_{m}=0.5 , \alpha=0.5, \nu = 1, \Lambda_{\text{eff}}=-1$ we find that $T'(r_h^{min})=0 \to r_h^{min} = 0.543748$ which corresponds to $T(r_h^{min})= 0.147635$.


The entropy of the black hole may be obtained using Wald's formula \cite{Wald:1993nt} which for our action reads
\begin{equation} S(r_h) = -2\pi\oint d^2x \sqrt{h} \left(\frac{\partial \mathcal{L}}{\partial R_{\alpha\beta\gamma\delta}}\right)\bigg|_{r=r_h}\hat{\epsilon}_{\alpha\beta}\hat{\epsilon}_{\gamma\delta}~,\end{equation}
where $\hat{\epsilon}_{\alpha\beta}$ the binormal to the horizon surface normalized to satisfy $\hat{\epsilon}_{\alpha\beta}\hat{\epsilon}^{\alpha\beta}=-2$ and $h$ is the induced metric on the horizon. Since the only quantity in the Lagrangian that involves the Riemann tensor is the Ricci scalar, we can obtain the standard Bekenstein-Hawking area law \cite{Brustein:2007jj}
\begin{equation} S(r_h) = 2\pi\mathcal{A}~,\end{equation}
where $\mathcal{A} = 4\pi [b_1(r_h)]^2$ is the area of the black hole. Hence
\begin{equation} S(r_h) = 8\pi^2 r_h(r_h+\nu)~,\end{equation}
with the scalar charge appearing in the entropy, resulting in higher entropy in comparison with the non-hairy black hole, since $\nu>0$.

To study the possibility of phase transitions, we will calculate the heat capacity. A positive heat capacity indicates that the black hole is thermodynamically stable. Non-stable black holes, may undergo a phase transition in order to be stabilized. Phase transitions occur at the points where the heat capacity vanishes or diverges. A vanishing point in the heat capacity indicates a first order phase transition, while a divergence point indicates a second order phase transition. The first order phase transition occurs at  high Gibbs
anergy and it does not change the favored configuration while a second order phase transition occurs at lower Gibbs anergy and allows the coexistence of two configurations.

\begin{figure}[h]
\centering
\includegraphics[width=.40\textwidth]{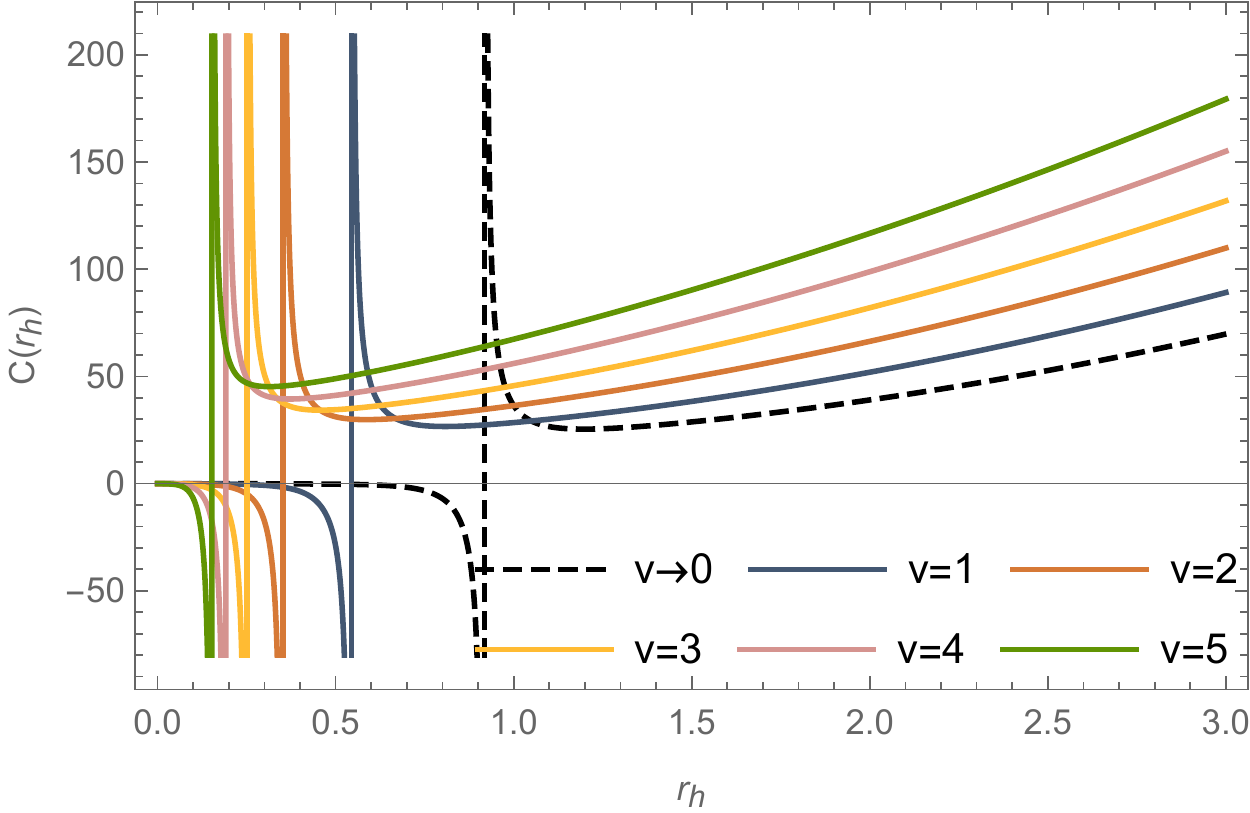}
\includegraphics[width=.40\textwidth]{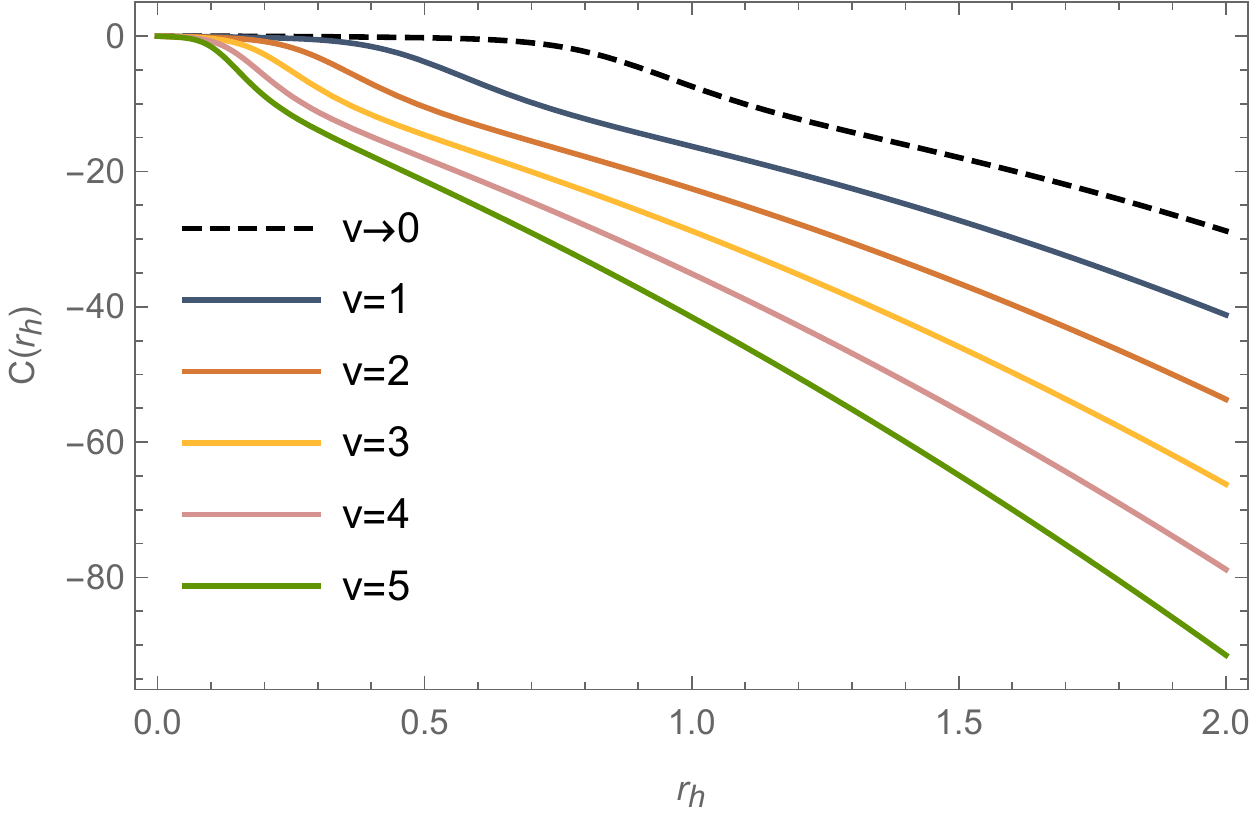}
\caption{The heat capacity for asymptotically AdS $\Lambda_{\text{eff}}=-1$ and flat $\Lambda_{\text{eff}}=0$ spacetimes, where we have fixed $Q_{m}=0.5 , {\color{blue}\alpha=0.5},$ while changing the scalar charge $\nu$.} \label{heat}
\end{figure}

The heat capacity is given
by
\begin{equation} C(r_h) = \frac{\partial m}{\partial T}\Big|_{r=r_h} =  \frac{m'(r_h)}{T'(r_h)}~,\end{equation}
where $m(r_h)$ is the mass as a function of the event horizon of the black hole, obtained from the relation $b(r_h)=0$.
The explicit expression is too complicated to be given here. For AdS spacetime, for large $r_h$, the heat capacity is positive, since
\begin{equation} C(r_h\gg1) \sim 2 \pi  r_h^2+2 \pi  \nu  r_h+\frac{\pi  \left(\nu ^2 \Lambda
   _{\text{eff}}-20\right)}{5 \Lambda
   _{\text{eff}}}+\mathcal{O}\left(\frac{1}{r_h^2}\right)~,\end{equation}
and the AdS black holes are stable. However, in order to see if the black hole undergoes a phase transition before it gets  stabilized we will plot the heat capacity in Fig. \ref{heat}. The fact is that the asymptotically AdS Euler-Heisenberg hairy black holes undergo a second order phase transition and then they are stabilized. The phase transition point occurs at the minima of temperature. For a flat spacetime, the heat capacity asymptotes to
\begin{equation} C(r_h\gg1) \sim -2 \pi  r_h^2-2 (\pi  \nu ) r_h+\mathcal{O}\left(\frac{1}{r_h^2}\right)~,
\end{equation}
where we can see that the flat black holes are thermodynamically unstable and there exists no phase transition to make the black holes stable as we can deduce from Fig. \ref{heat}.

\section{Energy Conditions}
\label{sec4}

\begin{figure}[h]
\centering
\includegraphics[width=.40\textwidth]{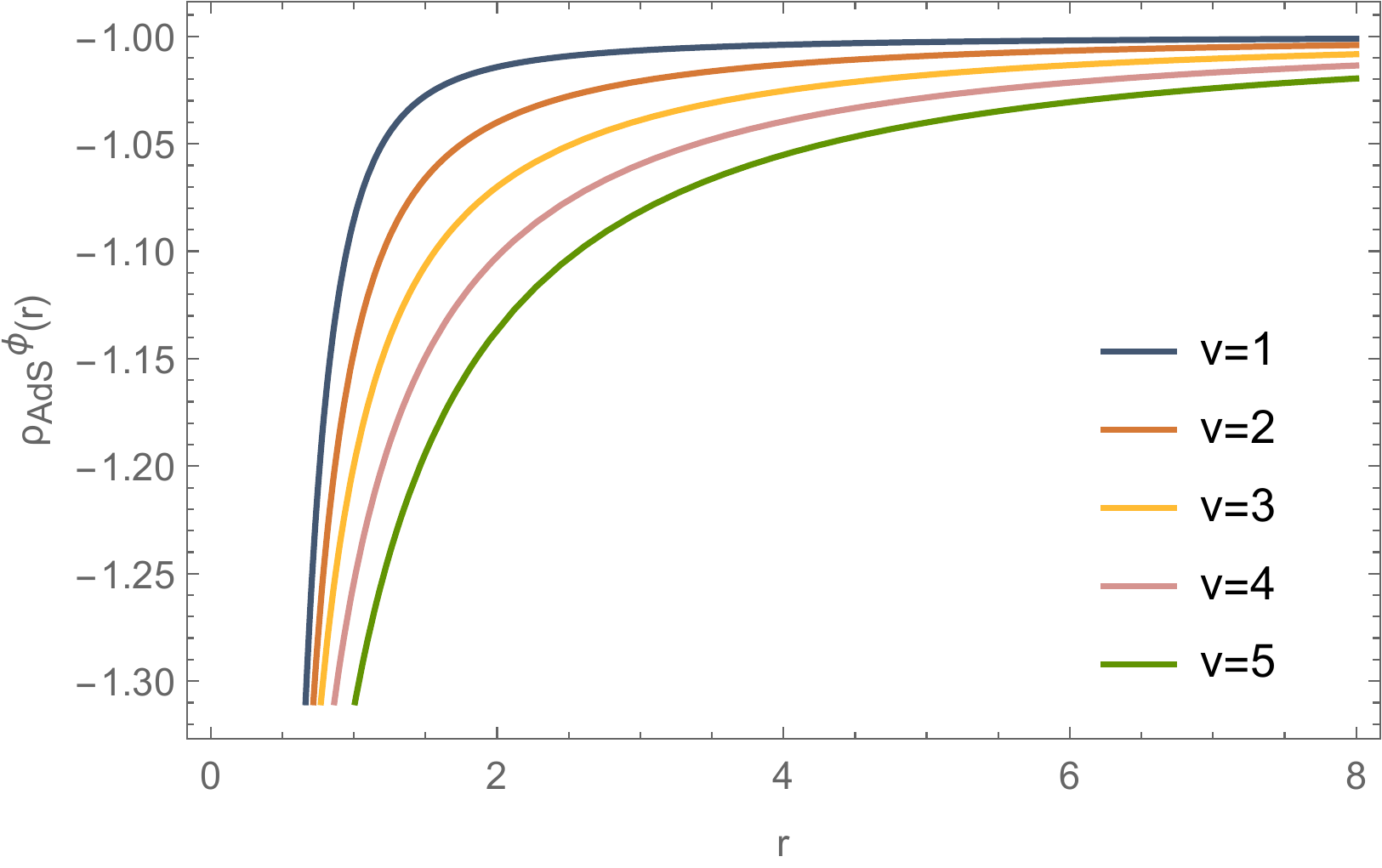}
\includegraphics[width=.40\textwidth]{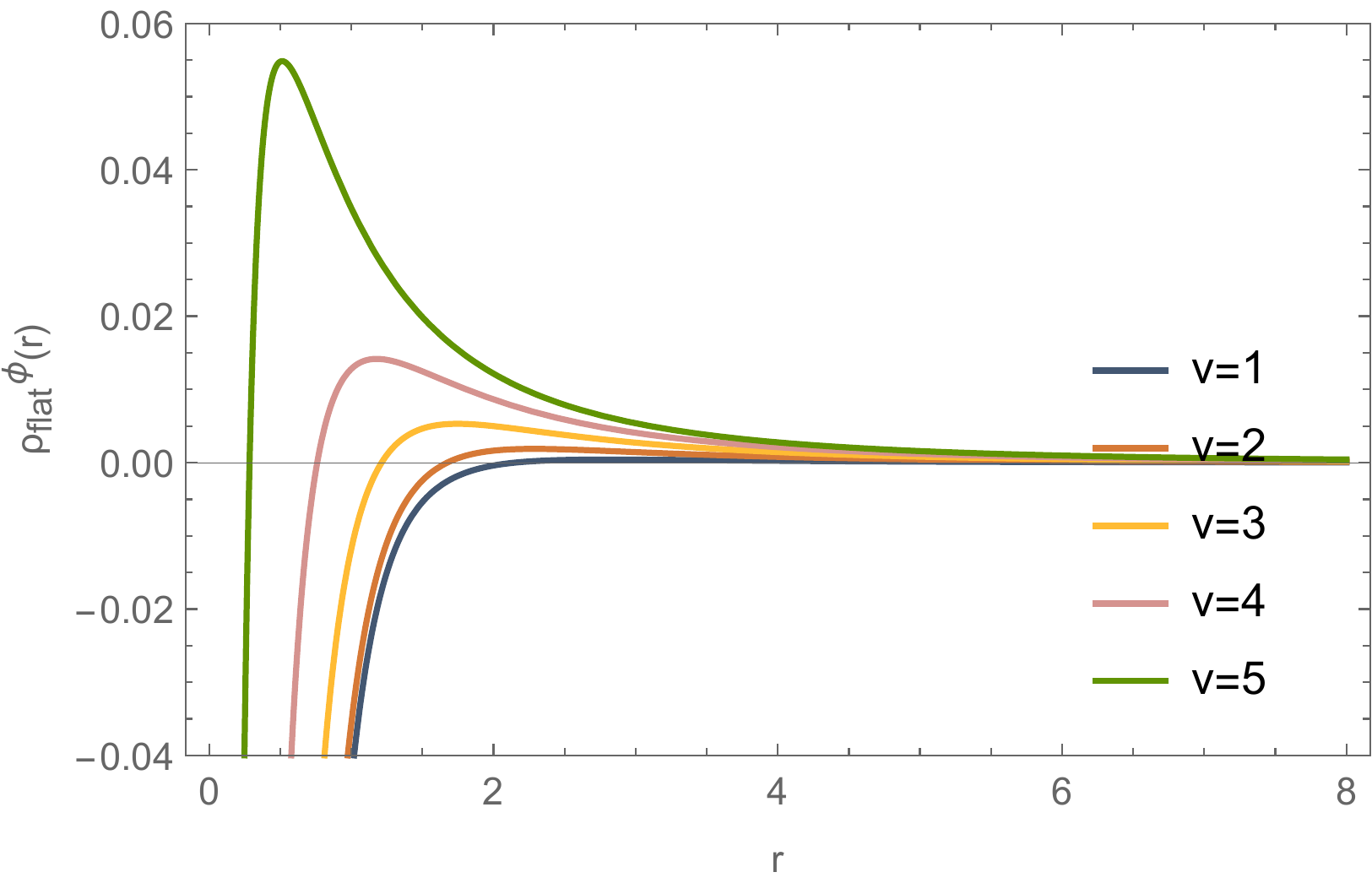}
\includegraphics[width=.40\textwidth]{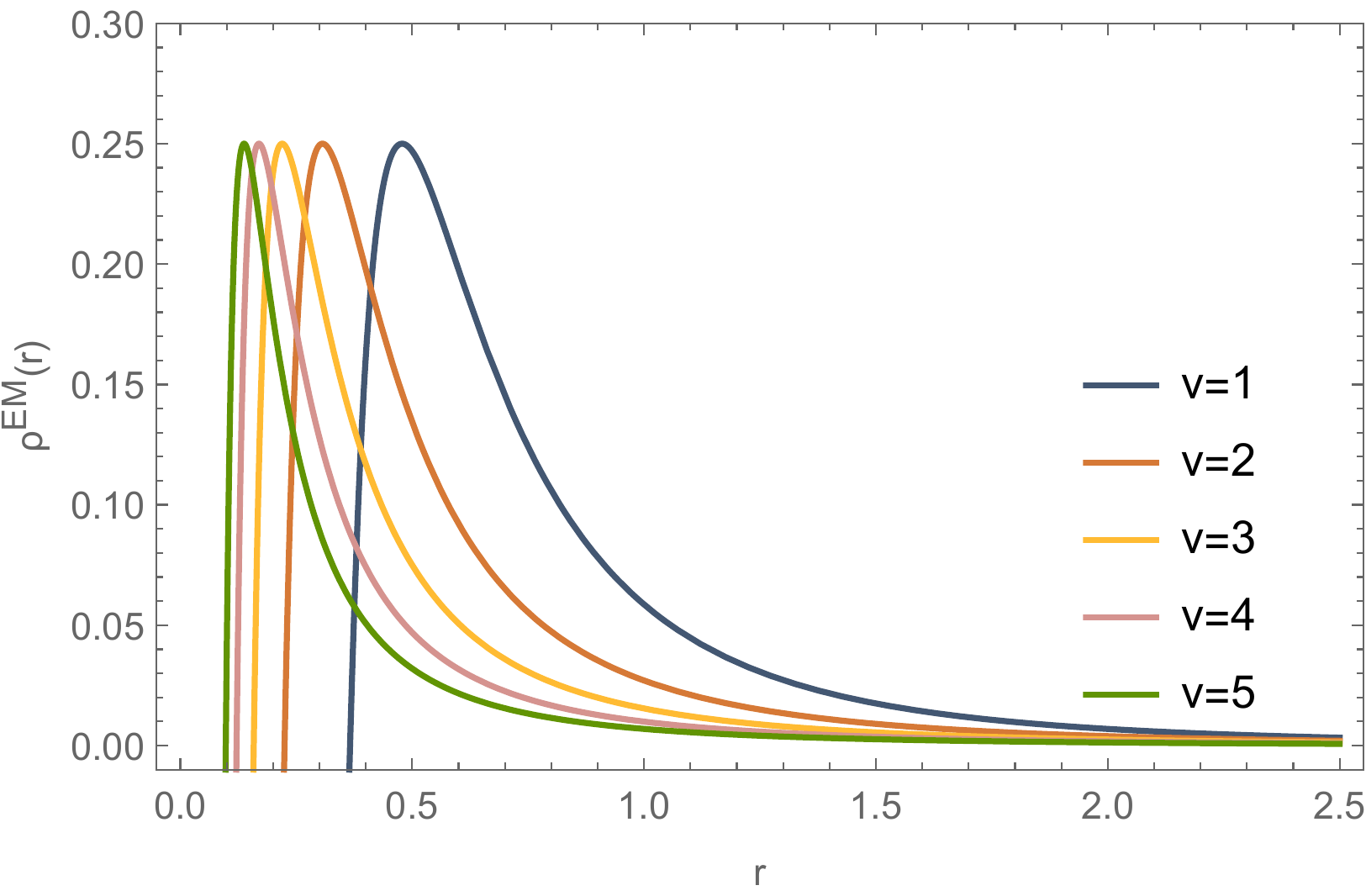}
\includegraphics[width=.40\textwidth]{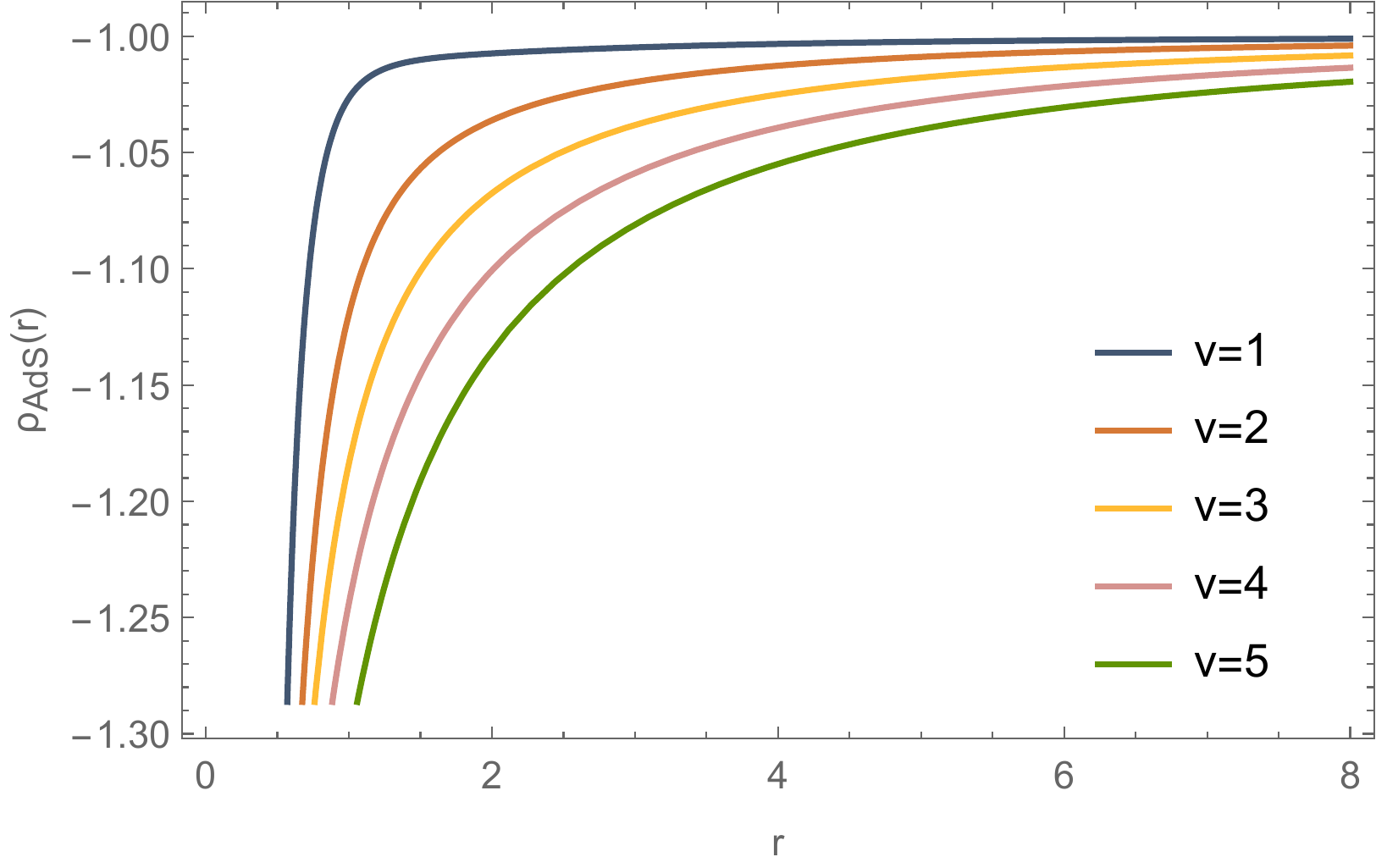}
\includegraphics[width=.40\textwidth]{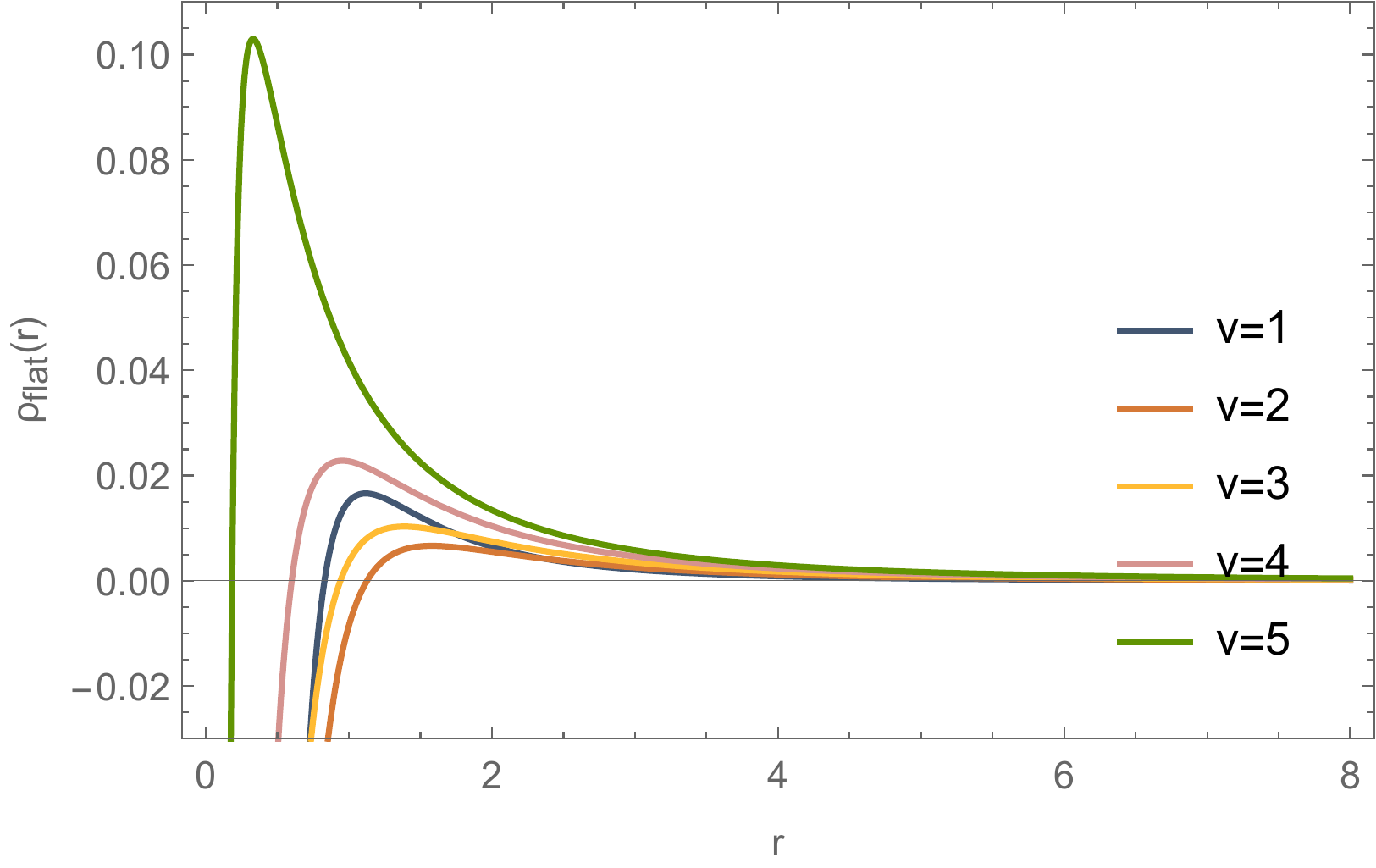}
\caption{Several energy densities are plotted. In the first row, we have the energy density of the scalar field for AdS (left) and flat spacetimes (right). In the second row, we plot the energy density of the electromagnetic part of the energy momentum tensor (left) and the total energy density for AdS spacetime (right). In the third row we plot the total energy density of the asymptotically flat case. For the AdS cases we have set $\Lambda_{\text{eff}}=-1$, for the flat spacetimes $\Lambda_{\text{eff}}=0$, while we have fixed $m=1, Q_{m}=0.5 , {\color{blue}\alpha=0.5}$, and we vary the scalar charge $\nu$.} \label{wec}
\end{figure}

In this Section we will discuss the energy conditions of the hairy black hole. For this reason, we will use the Einstein equation in the appropriate reference frame, where we can identify the energy density, the radial and tangential pressure as
\begin{eqnarray}
&&G^{\mu}_{~\nu} = T^{\mu}_{~\nu}~,\\
&& \rho = -T^{t}_{~t}~,\\
&&p_r = T^{r}_{~r}~,\\
&&p_{\theta} = p_{\varphi} = T^{\theta}_{~\theta}~.
\end{eqnarray}
The weak energy condition (WEC) states that given a timelike vector field $t^{a}$, the quantity $T_{ab}t^at^b$ is positive, i.e $T_{ab}t^at^b\ge0 \to \rho\ge0$. The null energy condition (NEC) states that $T_{ab}l^al^b\ge0 \to \rho + p_r>0$, where $l^al_a=0$, so that the geometry will have a focusing effect on null geodesics. For the energy momentum tensor of the scalar field, we have
\begin{eqnarray}
&&\rho^{\phi} = \frac{1}{2}b(r)\phi'(r)^2 +V(r) = -p_{\theta}^{\phi}~,\\
&&p_{r}^{\phi} =  \frac{1}{2}b(r)\phi'(r)^2 -V(r)~,
\end{eqnarray}
while for the energy momentum tensor of the Euler-Heisenberg theory we obtain
\begin{eqnarray}
&&\rho^{EM} =-\frac{2 \alpha  Q_m^4}{b_1(r)^8}+\frac{Q_m^2 \left(1-4 (\alpha -32
   \beta ) \mathcal{A}'(r)^2\right)}{b_1(r)^4}+\frac{4 \beta  Q_m
   \mathcal{A}'(r)}{b_1(r)^2}+6 \alpha  \mathcal{A}'(r)^4+\mathcal{A}'(r)^2 = -p_r^{EM}~,\\
&&p_{\theta}^{EM} = -\frac{6 \alpha  Q_m^4}{b_1(r)^8}+\frac{Q_m^2 \left(4 (\alpha -32
   \beta ) \mathcal{A}'(r)^2+1\right)}{b_1(r)^4}-\frac{4 \beta  Q_m
   \mathcal{A}'(r)}{b_1(r)^2}+2 \alpha  \mathcal{A}'(r)^4+\mathcal{A}'(r)^2~.
\end{eqnarray}
We will at first discuss the NEC, which implies $ \rho + p_r\ge0$. By adding the energy densities and radial pressures, we have
\begin{equation} \rho + p_r = \rho^{\phi} + \rho^{EM} + p_{r}^{\phi}+p_r^{EM} = \rho^{\phi} - p_r^{EM} + p_{r}^{\phi} +p_r^{EM} = b(r)\phi'(r)^2~.\end{equation}
First of all $\phi'(r)^2>0$ for any $r>0$. $b(r)$ is negative inside the black hole, resulting in the violation of the NEC, zero at the event horizon resulting to $\rho+p_r=0$, while after the event horizon $b(r)$ is positive, hence, the NEC is protected, regardless of the asymptotic nature of spacetime.
For the contribution of the scalar field to the total energy density, we can see that inside the event horizon, where $b(r)<0$, the WEC is violated by the scalar field, since $V(r)$ is also negative i.e $V(r)<0$ regardless of the asymptotic nature of spacetime. On the event horizon $b(r_h)=0$ and since $V(r_h)<0$ the WEC is also violated. Caution must be given for the contribution of the scalar field to the energy density in the causal region of the black hole i.e $r>r_h$. For asymptotically AdS spacetimes, outside of the event horizon we have $b(r)>0$ and $V(r)<0$, however, the scalar potential is too negative, hence as we can see in Fig. \ref{wec}, the scalar field part of the energy momentum tensor will always violate the WEC. For the asymptotically flat case, at large distances, the kinetic energy of the scalar field $\mathcal{T}(r) = b(r)\phi'(r)^2/2$ asymptotes to
\begin{equation} \mathcal{T}(r\to\infty) \sim \frac{\nu ^2}{4 r^4}+\frac{-5 \nu ^3-7 m \nu ^2}{10
   r^5}+\mathcal{O}\left(\left(\frac{1}{r}\right)^6\right)~, \label{kin} \end{equation}
while the potential behaves as
\begin{equation} V(r\to\infty) \sim -\frac{m \nu ^2}{5 r^5}+\mathcal{O}\left(\left(\frac{1}{r}\right)^6\right)~.\label{pot}\end{equation}
It is clear that their sum $\mathcal{T}(r) + V(r)$ will be positive at large distances, since the kinetic energy surpasses the contribution of the potential. It is therefore evident that for a region outside of the black hole horizon $\rho^{\phi}>0$.
The electromagnetic part of the energy density yields
\begin{equation} \rho^{EM}(r) = \frac{Q_m^2}{r^2 (\nu +r)^2}-\frac{2 \alpha  Q_m^4}{r^4 (\nu +r)^4}~. \end{equation}
There will be regions of negative energy density due to the Euler-Heisenberg modified electromagnetism parameter $\alpha$. We plot $\rho^{EM}(r)$ in Fig. \ref{wec} where we can see that $\rho^{EM}(r)$ can be positive, however it does not contribute much in the total energy density, hence $\rho<0$ everywhere and the WEC is violated in the case of  AdS spacetime. However in the asymptotically flat case, it is obvious from (\ref{kin}), (\ref{pot}) and FIG. \ref{wec}, that, for a region outside of the event horizon to infinity $\rho>0$ and the WEC holds.

\section{Conclusions}
\label{sec5}

We studied the Einstein-Euler-Heisenberg theory in the presence of a minimally coupled to gravity, self interacting scalar field. We solved analytically the field equations and, assuming an electromagnetic field with magnetic charge, we obtained novel magnetically charged hairy black holes. The scalar field dresses the black hole with secondary scalar hair,  while the scalar potential is negative in order to support the hairy structure and it possesses a mass term that satisfies the Breitenlohner-Friedman bound that ensures the perturbative stability of the AdS spacetime. The presence of  the scalar charge is introducing a new scale in the theory which leads to the appearance of  an effective cosmological constant. The hairy black hole develops three horizons when Euler-Heisenberg parameter and the magnetic charge $Q_m$ are  small and the horizon radius is getting large when the scalar charge and the gravitational mass are large.

We also studied the thermodynamics of the hairy  Euler-Heisenberg black hole. We found that the presence of matter outside the horizon of the black hole increases the temperature only for small black holes. Also we found the  same behaviour for the magnetic field, it increases the temperature only for small black holes. Calculating the heat capacity we found that  the asymptotically AdS Euler-Heisenberg hairy black hole undergoes a second order phase transition and then it is stabilized. The phase transition point occurs at the minimum  of the temperature while the scalar field gains entropy for the black hole by the addition of a linear term in the entropy and hence the hairy black holes are thermodynamically preferred.

We found that the WEC is violated on the horizon of the hairy  Euler-Heisenberg black hole. For asymptotically AdS spacetimes, outside of the event horizon the scalar field part of the energy momentum tensor will always violate the WEC. However in the asymptotically flat case, we found that for a region outside of the event horizon to infinity  the WEC holds.

It would be interesting  to extend this work to the case that the scalar field is magnetically charged. Then we expect that the
magnetized scalar field will interact with the magnetic field, so that  the magnetized scalar charge,  the magnetic charge and the Euler-Heisenberg parameter will play a decisive role in the structure and properties of the magnetized hairy Euler-Heisenberg black hole. It would also be of interest to study the shadow of the obtained spacetime and to constrain the modified Euler-Heisenberg parameter along with the scalar charge from the results of the Event Horizon Telescope \cite{EventHorizonTelescope:2019dse} in the astrophysical scenario $Q_m\ll m$. In \cite{Khodadi:2020jij} it was found that considering Maxwell electrodynamics there is a threshold value for the electric charge $Q$, above which any value of the scalar charge is allowed. It would be worth investigating the same possibility in our case.

\end{document}